\documentclass[fleqn,usenatbib]{mnras}

\usepackage[T1]{fontenc}
\usepackage{ae,aecompl}

\usepackage{graphicx}	
\usepackage{amsmath}	
\usepackage{amssymb}	

\usepackage{newtxtext,newtxmath}

\title[Chandra observations of nova KT Eridani in outburst]{Chandra observations of nova KT Eridani in outburst}

\author[Pei et al.]{
Songpeng Pei,$^{1,2}$\thanks{E-mail: songpeng.pei@student.unipd.it}
Marina Orio,$^{2,3}$
Jan-Uwe Ness$^{4}$
and Nataly Ospina$^{5,6}$
\\
$^{1}$Department of Physics and Astronomy, University of Padova, vicolo Osservatorio, 3, 35122
Padova, Italy\\
$^{2}$INAF-Osservatorio di Padova, vicolo Osservatorio, 5, 35122 Padova, Italy\\
$^{3}$Department of Astronomy, University of Wisconsin, 475 N. Charter Str., Madison, WI, USA\\
$^{4}$European Space Agency (ESA), European Space Astronomy Centre (ESAC), Camino Bajo del Castillo s/n, 28692 Villanueva de la Ca\~nada, Madrid, Spain\\
$^{5}$Department of Physics and Astronomy, University of Padova, Via Marzolo, 8, 35131 Padova, Italy\\
$^{6}$INFN Sezione di Padova, Via Marzolo, 8, 35131 Padova, Italy\\
}

\date{Accepted XXX. Received YYY; in original form ZZZ}

\pubyear{2015}

\begin{document}
\label{firstpage}
\pagerange{\pageref{firstpage}--\pageref{lastpage}}
\maketitle

\begin{abstract}
We analyse here four observations of nova KT Eri (Nova Eri 2009) done with the Chandra High Resolution Camera Spectrometer (HRC-S) and the Low Energy Transmission Grating (LETG) in 2010, from day 71 until day 159 after the
 optical maximum, in the luminous supersoft X-ray
 phase. 
 The spectrum presents many absorption features with a large range of
 velocity, from a few hundred km s$^{-1}$ to 3100
 km s$^{-1}$ in the same observation, and a few prominent
 emission features, generally redshifted by more than 2000 km s$^{-1}$.
 Although the uncertainty on the distance
 and the WD luminosity from the approximate fit
 do not let us rule out a larger absolute luminosity than our best estimate
 of $\simeq 5 \times 10^{37}$ erg s$^{-1}$, 
 it is likely that we observed only up to
 $\simeq$40\% of the surface of the white dwarf, which may 
have been partially hidden by clumpy
ejecta. Our fit with
 atmospheric models indicate a massive white dwarf in the 1.15-1.25 M$_\odot$ range.
 A thermal spectrum originating in the ejecta appears to be superimposed
 on the white dwarf spectrum. It is complex, has more than one component
 and may be due to a mixture of photoionized and shock ionized outflowing material.
 We confirm that the $\simeq$35 s oscillation that was reported earlier, was detected 
 in the last observation, done on day 159 of the outburst.

\end{abstract}




\section{Introduction} \label{sec:intro}

Novae in outburst are natural laboratories of extreme physical phenomena. They increase in optical luminosity from 8 to 17 magnitudes within hours, reach Eddington luminosity for a 1 M$_\odot$ star ($\simeq$10$^{38}$ erg s$^{-1}$), and emit copious flux at all wavelengths, from radio to gamma rays. 
A recent observational review is the one by \citet{2018acps.confE..57P}, while theoretical reviews can be found in \citet{2012BASI...40..419S, 2016PASP..128e1001S} and \citet{2005AIPC..797..319P}.
Nova eruptions are due to thermonuclear burning of hydrogen via the CNO cycle, at the bottom of a shell accreted by a white dwarf (WD) from a close binary companion. The outburst is repeated after periods ranging from a few years to hundreds of thousands years. Because the burning occurs in electron degenerate matter, it becomes explosive, inflating and possibly immediately ejecting part of the envelope. Since the initial suggestion of \citet{Bath1976}, it has been assumed that the bulk of the remaining envelope mass is then stripped by a radiation-pressure-driven wind, although a wind may also be triggered by Roche Lobe overflow \citep[e.g.,][]{2013ApJ...777..136W}. 

The evolutionary track of the post-nova is driven by a shift in the wavelength of the maximum energy toward shorter wavelengths, at a constant bolometric luminosity close to 10$^{38}$ erg s$^{-1}$ \citep[e.g.,][]{2012BASI...40..419S, 2013ApJ...777..136W}. Thus in X-rays, novae become some of the most luminous X-ray sources.
The supersoft X-ray source (SSS) appears during post-maximum optical decline and lasts in different novae from one week to $\approx$ 10 years as the WD photosphere shrinks back to pre-outburst radius, while thermonuclear burning still occurs near the surface. The WD reaches effective temperature T$_{\rm eff}>$250,000 K, so the energy peak is in the supersoft X-ray range. The SSS is indeed observed as predicted by the models \citep[e.g.,][]{Orio2012}.
The SSS wavelength range is easily absorbed and if the column density exceeds 10$^{22}$ cm$^{-2}$, the WD may never be detected. In most cases, however, the filling factor of the ejecta is sufficiently low that the WD becomes a luminous X-ray source, observable with X-ray gratings as far as the Magellanic Clouds.
By fitting atmospheric models abundances, effective temperature, white dwarf composition and mass can be derived.
Novae in the SSS phase also exhibit time variability, both periodic and aperiodic (e.g., \citealt{Ness2015} for the first, \citealt{Orio2018} for the latter).

Nova outflows also emit X-rays, usually with a thermal spectrum, but they rarely are luminous enough to obtain a grating spectrum with good S/N ratio. However, the emission lines of the ejecta may be superimposed on the spectrum of the central source, complicating the spectral fitting. In the nova we examine in this article, KT Eri, the central source appeared to be dominant and have a relatively ``clean'' spectrum without a very large contribution of the ejecta. 

Classical nova KT Eridani was discovered by \citet{2009CBET.2050....1I} at V=8.1 on November 25.536 UT (MJD 55160.536), 2009. In a pre-discovery study \citet{Hounsell2010} found that the maximum was on 2009 November 14.67 at unfiltered
 Solar Mass Ejection Imager (SMEI) magnitude 5.42$\pm$0.02 
 and V=5.4 \citep{Ootsuki2009}, after a pre-maximum halt for a few hours
 at 6th SMEI magnitude,
 with a pre-outburst rise of by 3 magnitudes in 1.6 days. The outburst amplitude was 9 mag and the estimated time for a decay by 2 magnitudes was 
t$_2$=6.6 days \citep{Hounsell2010}, indicating that KT Eri was a fast nova. KT Eri was spectroscopically classified as a ``He/N nova'', implying characteristics that are not the most frequent in novae \citep{Rudy2009}, and are more often detected in recurrent novae. The ejecta velocity reached a maximum of 3400 km s$^{-1}$ in H$\alpha$ 
\citep{Maehara2009}, and at later phases 2800$\pm$200 km s$^{-1}$ \citep{Ribeiro2011}.
The distance is $3.69_{-0.42}^{+0.53}$ kpc, obtained thanks to the {\sl Gaia} parallax\footnote{from the GAIA database using ARI's Gaia Services at \url{http://gaia.ari.uni-heidelberg.de/tap.html.}}.

\citet{Ribeiro2013} modeled the evolution of the H$\alpha$ profile, finding that the nova ejecta had the shape of a dumbbell structure with a ratio between the major to minor axis of 4:1, and an inclination angle of 
58$^{\circ +6}_{-7}$. From infrared observations, \citet{Banerjee2013} estimated an
 upper limit for the mass of the ejecta in the range 2.4--7.4 $\times 10^{-5}$ M$\odot$,
 but this range should be decreased by a factor of $\simeq$3
 to account for the fact that the authors assumed a larger distance than the GAIA limits imply.
 \citet{Drake2009} detected a variable
 progenitor and \citet{Jurdana2012} found that at quiescence the
variability had a 1 mag amplitude, with two distinct periods of 376 and 737 days. These authors also noted the similarity of the KT Eri outburst with those of known recurrent novae, but could not find any previous outburst in the Harvard plates and suggested that the outbursts may recur on timescales of few centuries. 
They also found evidence that the progenitor is very likely to be an evolved star, ascending towards the red giant branch.

KT Eri was first detected as an X-ray source in outburst with the {\sl Swift} XRT on day 39.8 after optical
 maximum, as a relatively hard source \citep{2010ATel.2392....1B}. It was still hard on day 47.5, but by day 55.4 a
 luminous SSS emerged. On day 65.6 the SSS was softened dramatically.
\citet{2010ATel.2392....1B} noted that the time-scale for the SSS turn-on was very similar in the recurrent nova LMC 2009a.
 The timing analysis of the {\sl Swift XRT} data from day 66.60 to 79.25 revealed a 35 s modulation \citep{Beardmore2010}, that was later confirmed by \citet{Ness2014, Ness2015} with {\sl Chandra}. \citet{Ness2015} showed that this modulation was detected only during the early SSS phase, and then again much later, on day 159. With the {\sl Chandra} LETG, the nova X-ray spectrum 
 had several similarities with that of N SMC 2016 \citep{Orio2018}, with the luminous continuum and deep absorption lines of many novae X-ray spectra \citep{Ness2010}.

\section{Observation and data reduction} \label{sec:observation}
KT Eri was observed by the HRC-S and the LETG of Chandra four times in 2010.
 The details of the observations are shown in Table~\ref{table:obs}. 
 The principal investigator of the first observation was Jan-Uwe Ness \citep{Ness2010}, and Jeremy
 Drake was the principal investigator of the three following ones.
The net exposure time for the first observation was about 15 ks, 
while the other three observations were shorter: the 
nominal exposures were to be of 5 ks, but some events could not be 
telemetered and the dead time corrected exposures were only of about 3 ks.

 The {\sl Chandra}
X-ray data analysis software CIAO v4.11 and calibration package CALDB 4.8.5 were used for the data reduction. The background subtracted light curves measured
 with the Chandra HRC-S camera (0 order) were extracted from event files with “DMEXTRACT”. 
The script “chandra$_{-}$repro” was used to extracted the Chandra HRC-S+LETG high resolution spectra, and the
script “combine$_{-}$grating$_{-}$spectra” was used to combine the first order grating redistribution matrix files (rmf), ancillary response files (arf) and the grating spectra. 
\begin{table*}
\caption{Chandra observations of KT Eridani examined in
 this article, and measured count rates for the X-ray detectors. The LETG count rate is for the summed +1 and -1
 orders of LETG camera, and the HRC-S count rate is for the 0 order of HRC-S camera. 
The X-ray flux is measured between 18 and 80 \AA \ by integrating the flux measured
 with the LETG.
}
\label{table:obs}
\begin{center}
\begin{tabular}{rrrrrrrr}\hline\hline \noalign{\smallskip}

 Instrument & Date$^a$ & ObsID & Day$^b$ &  Exp. time$^c$ & C.R. (LETG)  & C.R. (HRC-S) & F$_{\rm x}$  $\times 10^{-9}$  \\
            &  (UTC)     &           &          & (ks)         & (counts s$^{-1}$) & (counts s$^{-1}$) &
 (erg s$^{-1}$ cm$^{-2}$) \\

\hline \noalign{\smallskip}
 Chandra HRC-S+LETG & 2010 Jan 23     & 12097 & 71.3 & 14.950  & 19.39$\pm$0.04  & 13.87 & 0.95 \\
 Chandra HRC-S+LETG & 2010 Jan 31     & 12100 & 79.3 & 2.753  & 111.70$\pm$0.24  & 84.85  & 5.94 \\
 Chandra HRC-S+LETG & 2010 Feb 06     & 12101 & 84.6 & 3.520  & 74.48$\pm$0.18  & 58.20 & 3.89  \\
 Chandra HRC-S+LETG & 2010 Apr 21     & 12203 & 158.8 & 3.206  & 130.30$\pm$0.23  & 113.74 & 7.89 \\
\hline \noalign{\smallskip}
\end{tabular}
\end{center}
Notes:\hspace{0.1cm} $^a $: 
Start date of the observation.
 $^b $: Time in days after the optical-maximum on 2009-11-13. $^c $: Exposure time of the observation (dead time corrected).
\end{table*}
\section{Timing analysis} \label{sec:timing}
\begin{figure*}
\centering
\includegraphics[width=17.0cm]{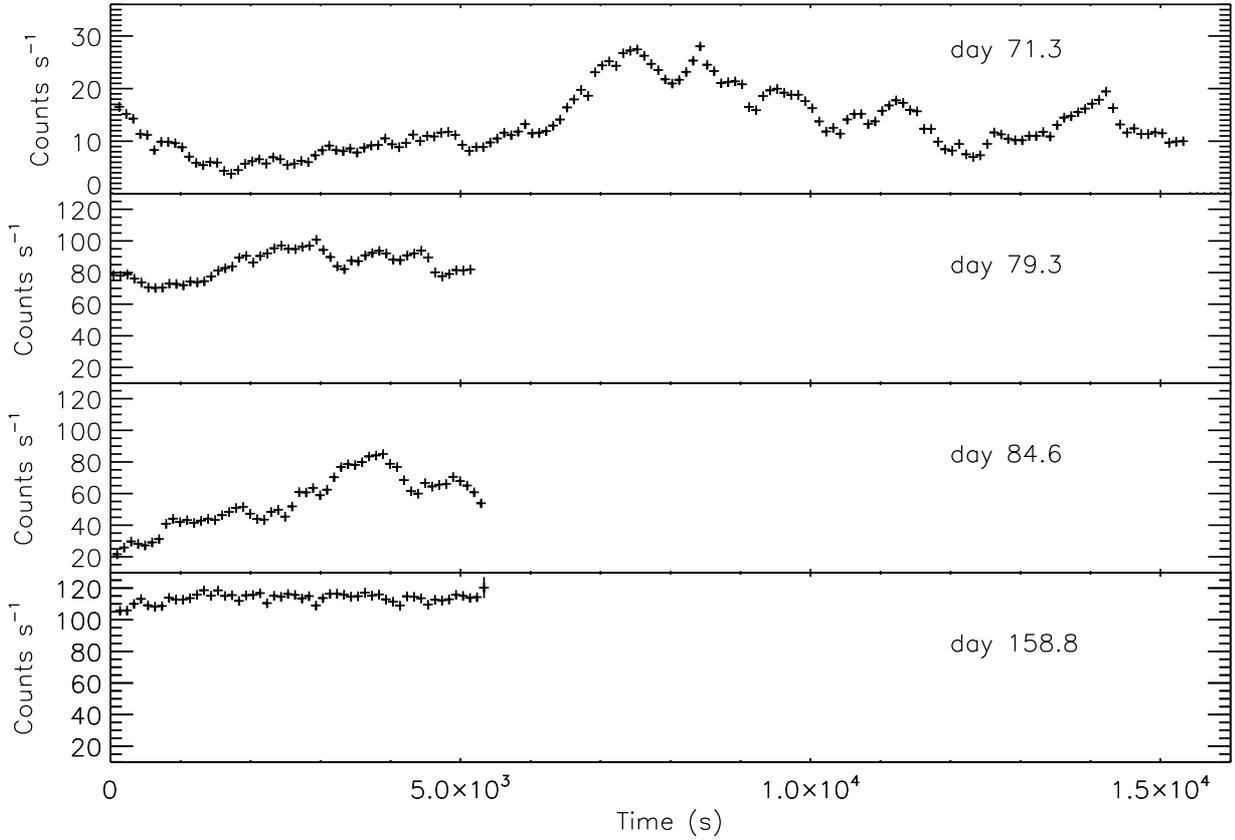}
\caption{The background subtracted zero order light curves of KT Eridani measured with the HRC-S camera
on days 71.3, 79.3, 84.6 and 158.8, binned every 100 s.
 Note that a different scale has been used in the y axis for the first exposure.
 The HRC-S camera is calibrated in the 0.06-10 keV range,
 but we know from the LETG spectrum that the source signal was above the
 background only between 155 and 620 eV.}\label{fig:lc}
\end{figure*}

\begin{figure}
\begin{center} 
\includegraphics[width=0.98\columnwidth]{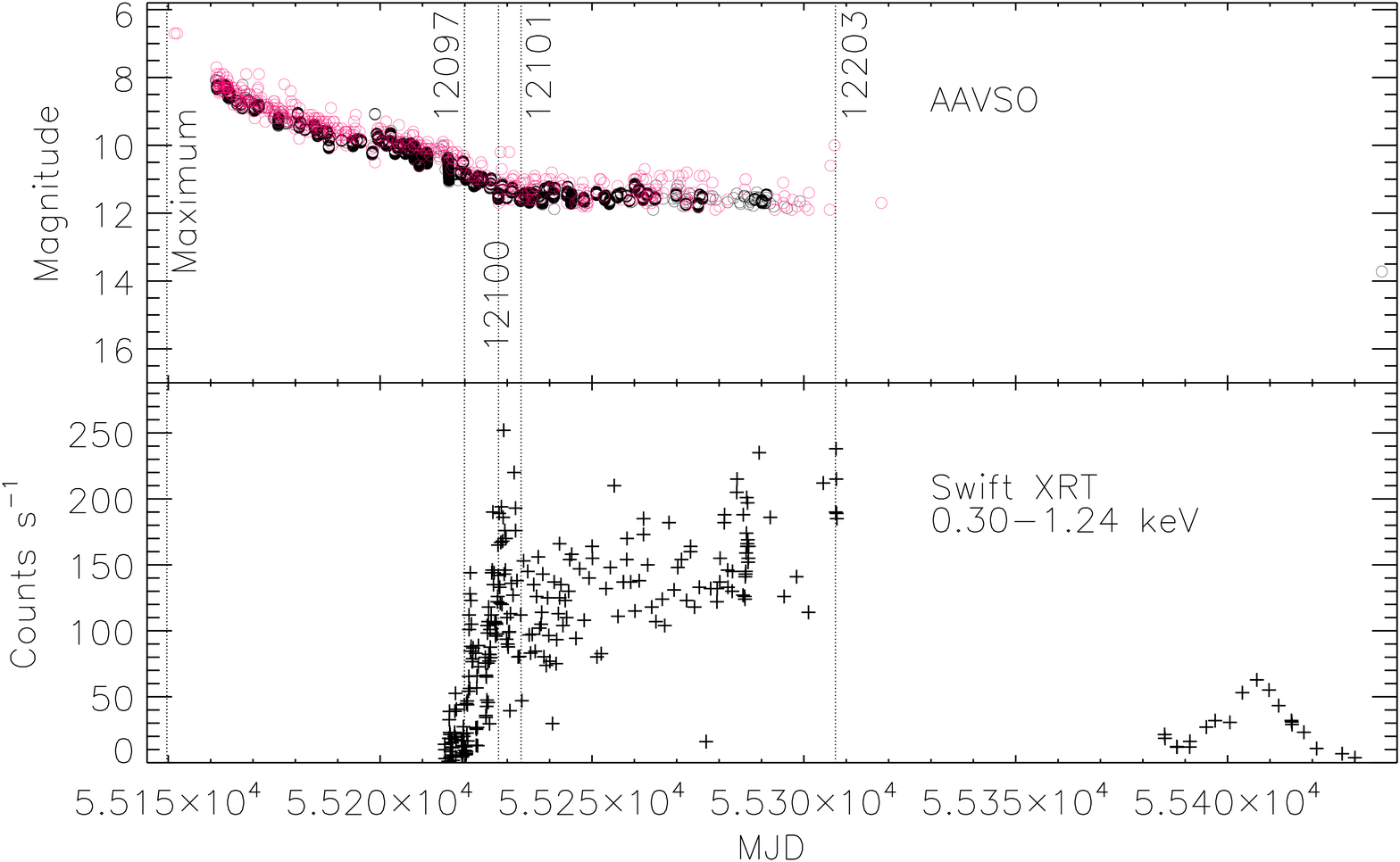}
\caption{The top panel shows the 
AAVSO visual light curve of KT Eridani (red) and its AAVSO light curve in the V band (black).
 The first vertical line marks the maximum of the outburst 
in the optical band (MJD 55149.67), and the others indicate the times
 of the observations done with {\sl Chandra}.
In the lower panel we show the {\sl Swift} XRT light curve in
 the 0.30-1.24 keV band (corresponding
 to 10–41 \AA).}
\label{fig:sum}
\end{center}
\end{figure}

The background subtracted zero order light curves measured with the HRC-S camera, binned every 20 s for visual purposes, 
are shown in Fig.~\ref{fig:lc}. 
 The nova was also monitored with {\sl Swift} and in Fig.~\ref{fig:sum}
 we show the {\sl Swift} X-Ray Telescope (XRT) light curve and
 the AAVSO optical light curve in the visual and V band.
 The {\sl Swift} XRT exposures were typically about 1000 s long.

The count rates vary significantly
during all the four {\sl Chandra} observations on relatively short time scales, 
and the average
 count rate also varied in the different exposures.

On day 71.3, the
count rate varied by a factor of 7.8 during the exposure.
 On day 79.3 the variation was by a factor of 1.6, on day 84.6 by a factor of 3.9, on day 158.8 by a factor of 1.4.
\subsection{Appearance and disappearance of a short periodic modulation}
 A period of 35 s was detected in the light curve of Swift XRT
 data by \citet{Beardmore2010}. This period was only detected on the
 {\sl Chandra} observation of day 158.8 by \citet{Ness2015}. These
 authors calculated power spectra
with the method of \citet{1986ApJ...302..757H}.

\begin{figure*}
\begin{center}
\includegraphics[width=88mm]{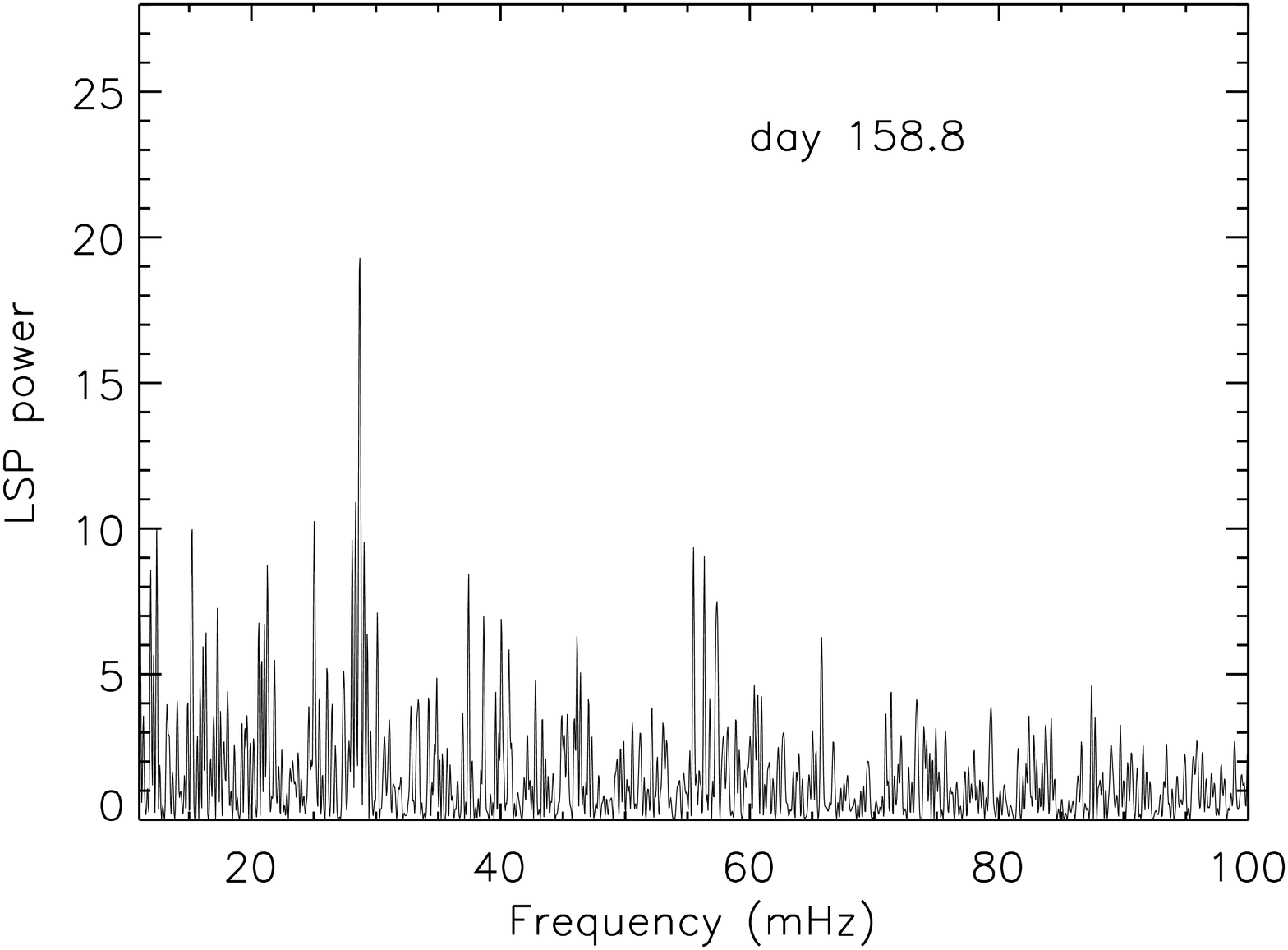}
\hspace{-0.62em}
\includegraphics[width=88mm]{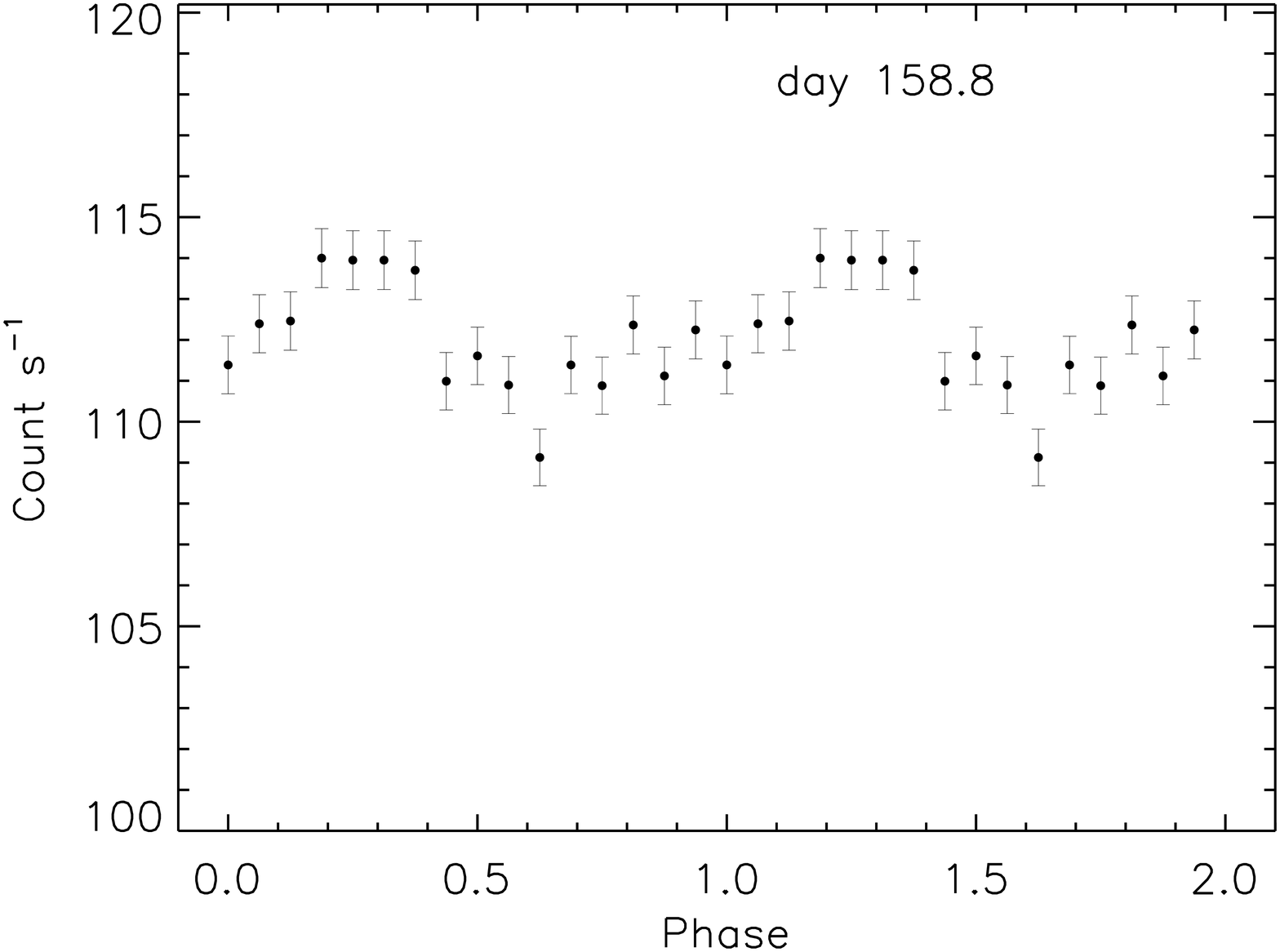}

\caption{Left: Lomb-Scargle periodogram of the light curve of the fourth observation 12203 of KT Eridani. Right: the light curve folded with the detected period 34.83 s (corresponding to $\simeq$ 28.71 mHz). The epoch of the first data point is the zero point.}
\label{fig: lsp and pfold lc}
\end{center}
\end{figure*}

We confirm the detection, having found a period of $34.83 \pm 0.06$ s with 99.0\% significance
in the day 158.8 light curve. We used the Lomb–Scargle method \citep{1982ApJ...263..835S}. We first detrended the light curves by subtracting the mean and dividing by 
the standard deviation.
The Lomb-Scargle periodogram (LSP) was calculated
 by us with the Starlink PERIOD package. 
The PERIOD task SCARGLE was used to create a LSP of each light curve.
 The PERIOD task SCARGLE uses the Lomb–Scargle method by scanning 
in the frequency space.
With SCARGLE we obtained the LSP frequency plot
 shown in the left panel of Fig.~\ref{fig: lsp and pfold lc}.
 With the PERIOD task PEAKS we found 
the highest peak in the periodogram between the frequencies we specified, and in order to determine
 the statistical significance of the
 period and obtain a statistical error, we performed a Fisher randomization test,
 as described by \citet{Nemec1985}, over frequencies from 10 to 100 mHz, including
 also red noise in the significance analysis.
The period was only detected in the light curve of the fourth observation.
 Considering
 that this period may have been missed if it was only present for a short time during an 
exposure,
 we divided the light curves of the other three observations into 
 segments of 1000 s. A
 period of $\simeq$35 s period was then detected in two
 segments of the first exposure, the eighth and the fourteenth in
 our subdivision in 1000 s intervals. However,
 the corresponding peaks in the periodogram are not very
 prominent. The results are shown in Table~\ref{tab:Period}.
 No periodicity was retrieved in
 any of the intervals of the second and third exposure.
 Fig.~\ref{fig: lsp and pfold lc} shows the LSP of the light curve on day 158.8 extracted with DMEXTRACT without any energy filtering and the same light curve folded on the corresponding detected period shown in Table~\ref{tab:Period}.

\begin{table}
\begin{minipage}{80mm}
\caption{Temporal analysis of the light curves showing the $\sim$35 s period.}
\label{tab:Period}
\begin{tabular}{lccc}
\hline
\hline

      Observation & Period & Significance & Modulation   \\    
      (or segment) & (s) & (\%) & amplitude$^a$ (\%)                    \\
\hline

day 71.3$^b$ (08)$^c$ &$34.49 \pm 0.30$  & 67.5  &4.52   \\
day 71.3$^b$ (14)$^d$ &$35.29 \pm 0.30$  &85.5  &9.68  \\
day 158.8$^b$ &$34.83 \pm 0.06$   &99.0  &2.18 \\

\hline


\end{tabular}
Notes:\hspace{0.1cm} $^a $: We define the period modulation amplitude as $(max-min)/(max+min)$.
 $^b $: Time in days after the optical-maximum on 2009-11-13. $^c $: Segment 8 in
 the exposure of day 71.3. $^d $: Segment 14 in the exposure of day 71.3.
\end{minipage} 
\end{table}

We did not detect any other periodicities in all four observations. 
The proposed periods of 0.09381 d (or 8105.18 s, Taichi Kato, vsnet-alert 
11755\footnote{\url{http://ooruri.kusastro.kyoto-u.ac.jp/mailarchive/vsnet-alert/11755.}})
 and $0.1952 \pm 0.0013$ d ($16865.28 \pm 112.32$ s, \citealt{2018NewA...58...53B}),
detected in the optical
light curves, are too long to be measured in the {\sl Chandra} exposures.

\begin{figure*}
\centering
\includegraphics[width=17.0cm]{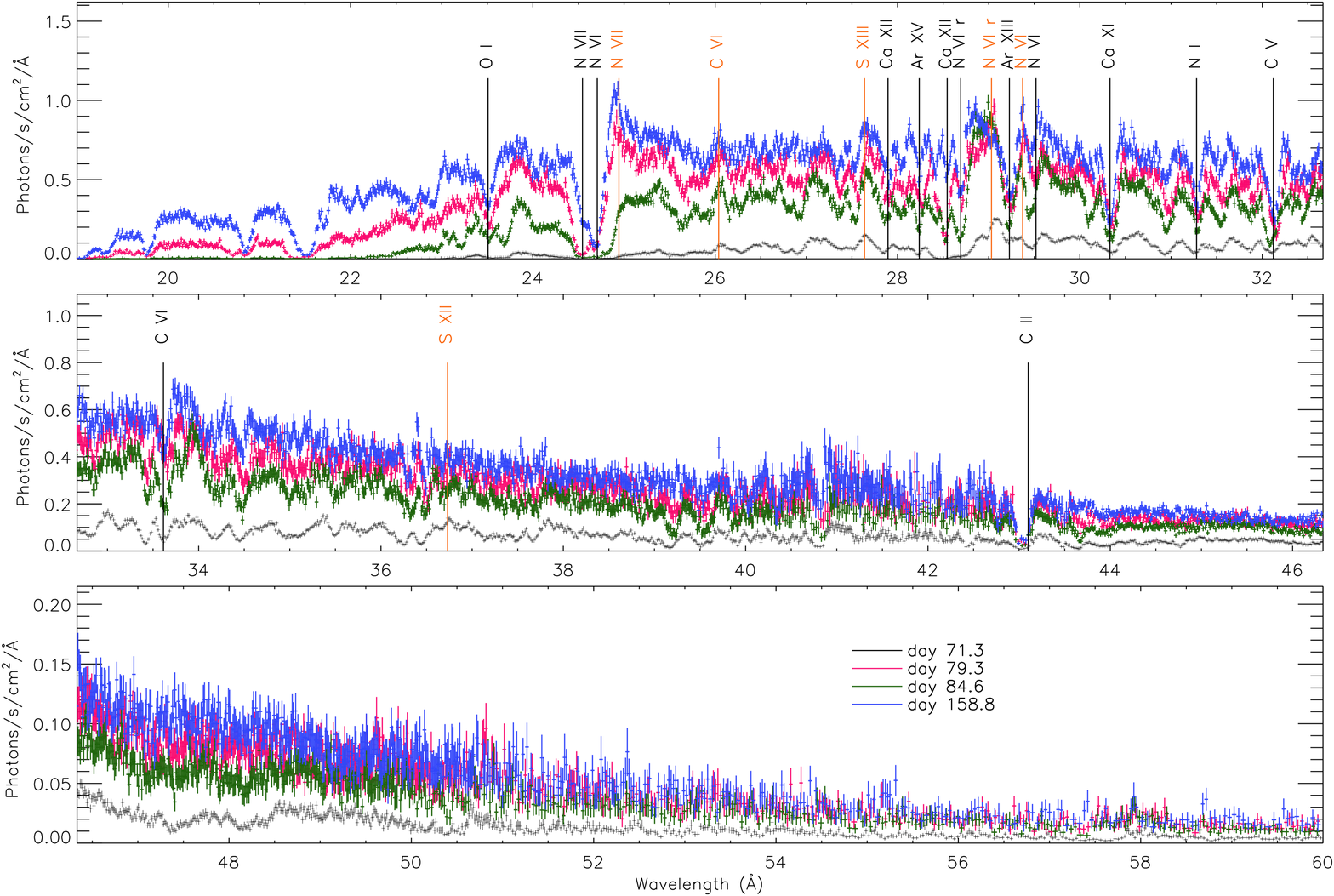}
\caption{The spectra of KT Eridani on days 71.3 (black), 79.3 (pink), 84.6 (green) and 158.8 (blue) after the optical-maximum. The absorption lines identified
 for the day 158.8 spectrum are marked
 in black and the strong emission lines are marked in orange, with the proposed identifications.
The spectra were binned for better clarity, and the spectra on days 71.3, 79.3, 84.6 and 158.8 were binned with at least 20, 40, 35 and 45 counts per bin, respectively.}\label{fig:4spectra}
\end{figure*}
\section{The high resolution X-ray spectra}
In Fig.~\ref{fig:4spectra} we show the summed +1 and -1 order LETG spectra
 of the four observations, and we indicate a number of clearly
 detected absorption and emission lines; these are the strongest ones.

\begin{figure*}
\centering
\includegraphics[width=8.00cm]{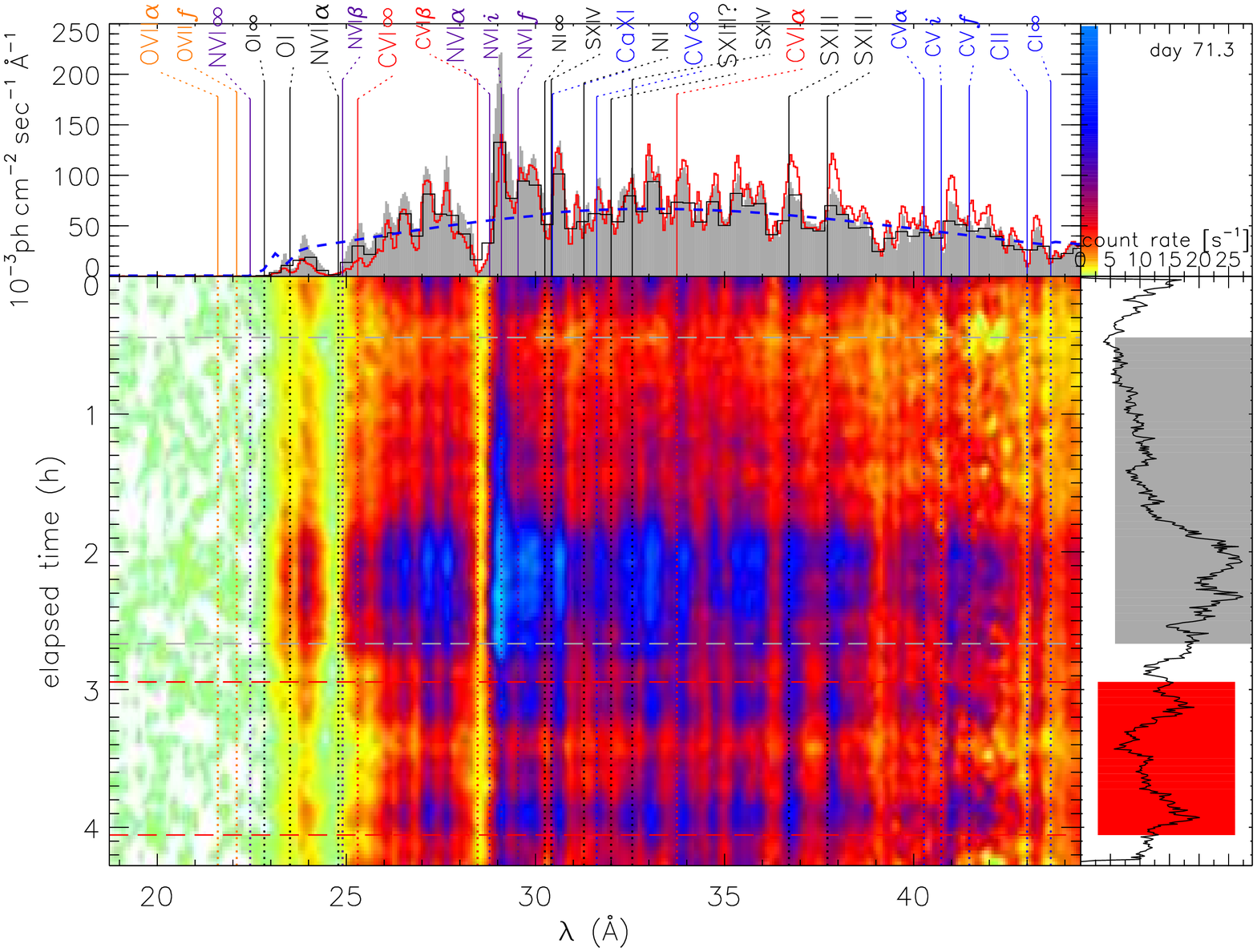}
\hspace{1.62em}
\includegraphics[width=8.00cm]{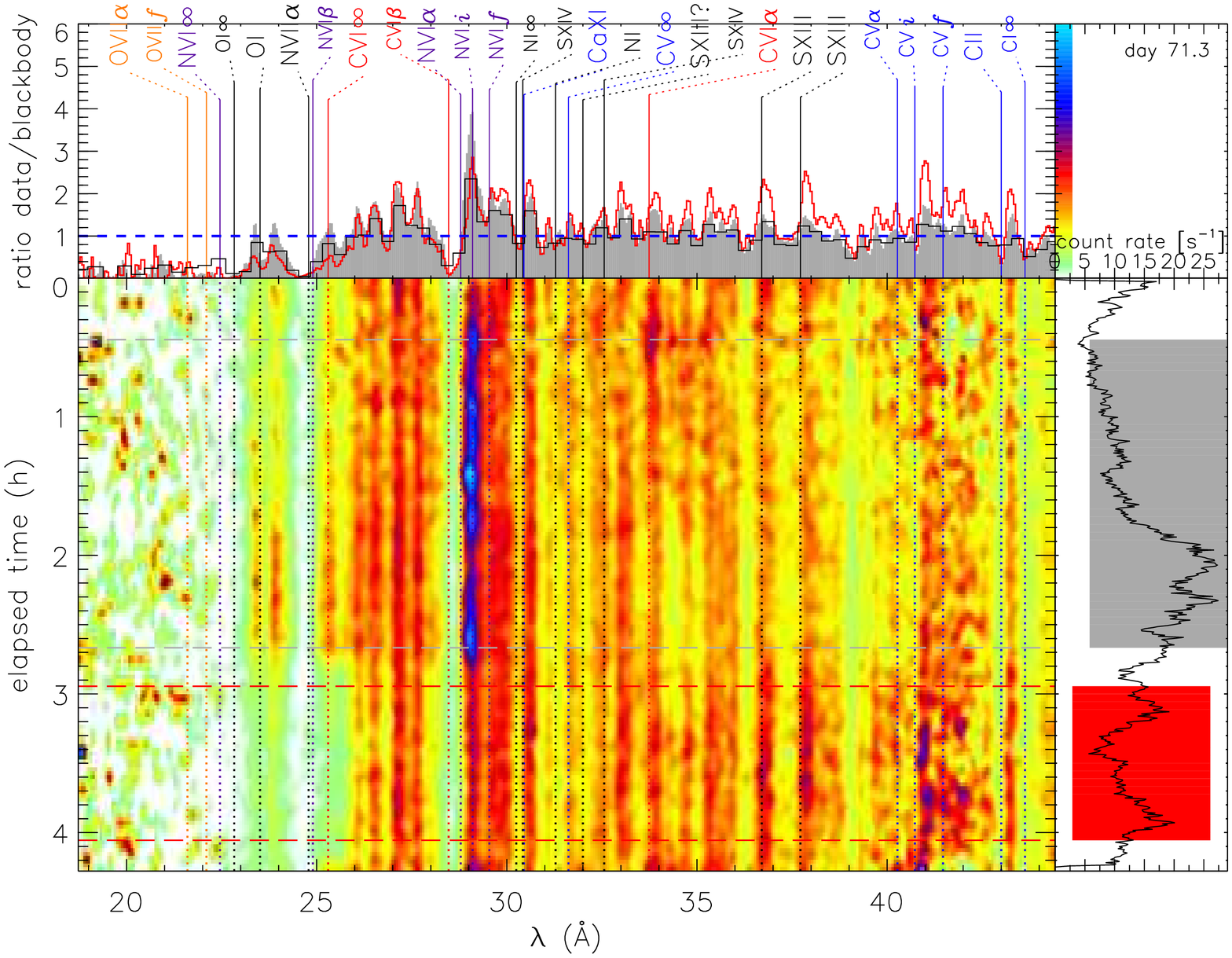}
\hspace{1.62em}
\includegraphics[width=8.00cm]{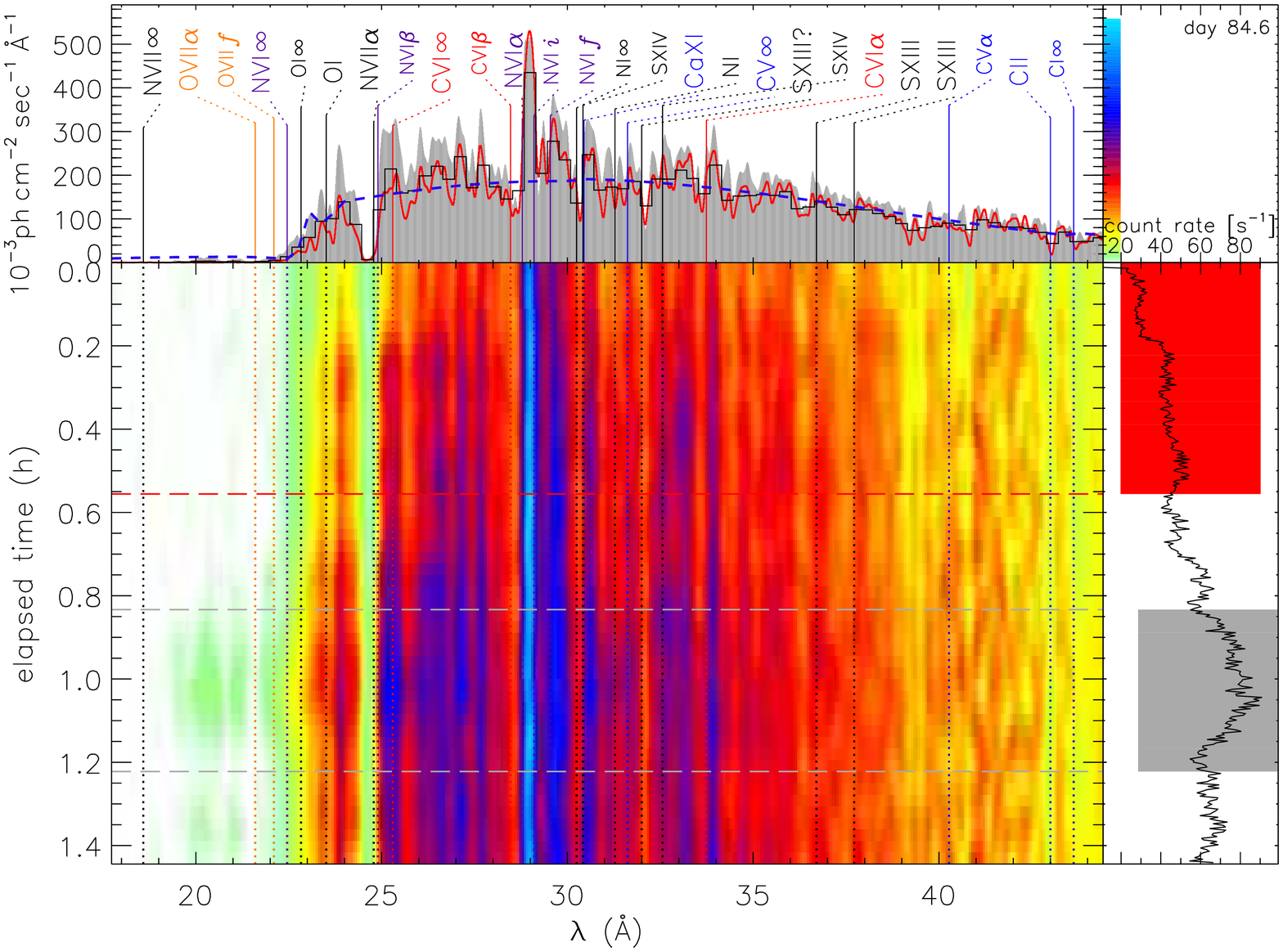}
\hspace{1.62em}
\includegraphics[width=8.00cm]{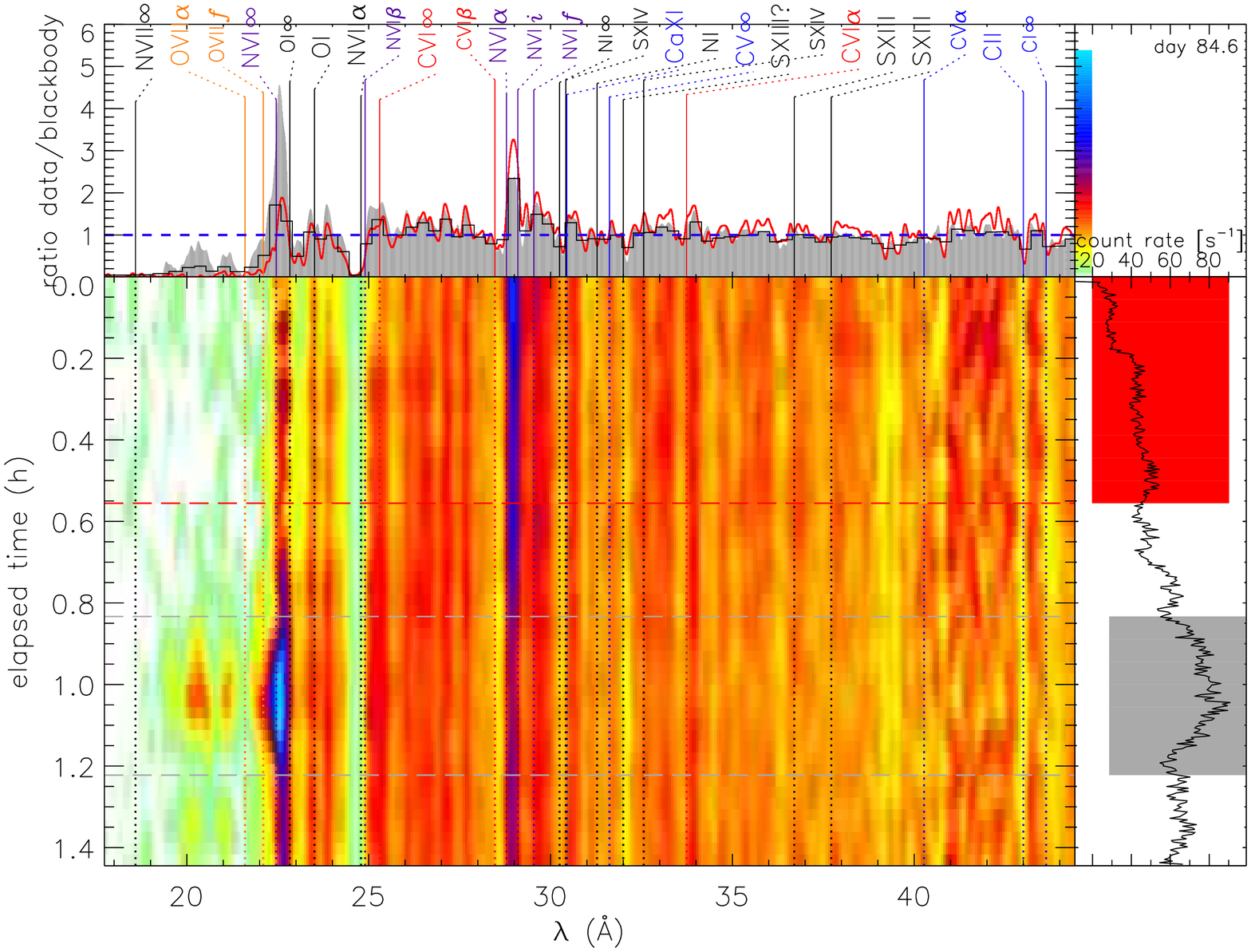}
\caption{
Spectral Time Maps illustrating the spectral evolution. All the important
 atomic features are indicated on the top left panel (for
 reference, and independently of actual
detection). The day
of observation is given in the top right corners (day 71.3 top
row and day 84.6 bottom row). Each plot is divided into 4 panels: Top
left: photon flux versus wavelength. The red line and the grey shades
show the spectra extracted of selected time intervals and the colours correspond to
those in the bottom right panel, where the time intervals 
are marked. 
The blue dashed line shows the best blackbody fit for the
continuum. The bottom left, larger plot shows
 the colour-coded intensity spectral
map as a function of time, in units of hours. The
 time since the beginning of the exposure 
is ordinate, growing towards the bottom and the wavelength is in the abscissa.
 The colours represent photon fluxes; 
the legend key is shown close to the vertical photon flux axis in the top
right panel. Each of the bottom right panels shows the zero-order light curve
 (repeating Fig.~\ref{fig:lc}) with count rate as ordinate 
and time as abscissa, and the time axis is shared with the spectral map panel. 
 The shaded areas marking time interval from which spectra have been
extracted that are shown in the top left panel.
 The figures on the right show time evolution maps of the spectra normalized
by the result of the blackbody fit, with $T_{\rm blackbody}$=570,000 K,
N(H)=5.6 $\times 10^{20}$ cm$^{-2}$, oxygen abundance 25 times the solar value in
 the intervening interstellar or circumstellar medium (ISM, CSM) on day 71.3
 and $T_{\rm blackbody}$=740,000 K,
N(H)=5.8 $\times 10^{20}$ cm$^{-2}$, oxygen abundance 16 times solar value in
 the intervening ISM or CSM on day 84.6. This approximate 
fit allows the deviations from the continuum to be more easily identified. 
}\label{fig:smap}
\end{figure*}

Fig.~\ref{fig:smap} illustrates the spectral evolution 
 during the two exposures with large flux variations, which also
 presented spectral variations with the flux, namely 
day 71.3 (top row) and day 84.6 (bottom row). The
respective plots on the left are the observed spectra, and on the
right we show the time maps normalized to a blackbody curve that approximately
 fits the continuum. We first found the 
best fit blackbody curve for each observation, shown in the top
left panel of the original time map with the dashed blue line, 
and following the blackbody we adjusted the normalization for the respective
spectra.
In the central panel for day 71.3 the N\,{\sc vi} line at
28.79\,\AA\ grows after $\sim 0.5$ hours, while the zero-order light
curve indicates this happened during an episode of still rather low
count rate, before a rise. 
In the blackbody-normalized time
map, this emission line appears more constant, indicating that it has
grown together with the underlying continuum. However, $\sim 2.8$
hours after the beginning of the exposure, this line has shrunk considerably more
than the continuum.
 We also note that at the same time, the continuum
shortward of $\sim 26$\,\AA\ (possibly related to the C\,{\sc vi}
ionization edge at 25.3\,\AA) has decreased notably while some emission
lines at long wavelengths (e.g., 31.8\,\AA, 32.8\,\AA, 41\,\AA, possibly
S\,{\sc xiii} and C\,{\sc v}) have increased in strength.
Thus, the emission
lines above the continuum have thus not evolved in correlation or
anti-correlation with the strength of the continuum.

Similarly, on day 84.6, the evolution of emission lines is not aligned with that
 of the
continuum. During the brightest part of the light curve,
$\sim 1$ hour after the beginning of the exposure (grey-shaded area in the
bottom right panel), the N\,{\sc vi} line at 28.78\,\AA\ has become
slightly weaker relative to the continuum (thus has not increased in flux as much as the continuum) while shortward of $\sim$ 23\,\AA\ (possibly related to
O\,{\sc i} ionization edge in the cool gas), a few emission features have grown
substantially. Possibly, the higher count rate was caused by a temporary
reduction of the neutral oxygen column density. The surrounding cool
material may be inhomogeneous, or it may have been temporarily ionized
which would make it more transparent to high-energy radiation.
\section{The individual lines: identification, fluxes, blue and red shift}
We found absorption lines due to transitions of nitrogen, carbon, calcium and argon
in all four spectra.
The method described by \citet{2010AN....331..179N, 2011ApJ...733...70N} was used to determine the line shifts, widths,
and optical depths at the line center of the absorption lines for the spectra for the
 four exposures. Following \citet{2011ApJ...733...70N}, we did
 not include the absorption correction, because it is not important in determining
 velocity and optical depth.
 The narrow spectral region around each line is fitted with a function
 \begin{equation}
  C(\lambda) \times e^{-\tau(\lambda)}
 \end{equation}.

Following \citet{2011ApJ...733...70N}, the optical depth $\tau(\lambda)$ is defined as the Equation (2) in \citet{2010AN....331..179N}, and $\lambda_{\rm c}$ and $\sigma$ in this equation are the central wavelength and the Gaussian line width which is always
 larger than the instrumental broadening, respectively. We assumed $C(\lambda)$ is a linear function for each line in modelling
 the continuum. We did compare the result with a continuum modelled
 with an absorbed black-body model and an absorbed TMAP model,
 and found that the model chosen for the continuum in the small
 wavelength range subtended by each line does not change
 the results. The Gaussian width of each line in velocity space was computed as 
$v_{\rm width}=\sqrt{ ({\rm c}\sigma/\lambda_0)^2-v_{\rm instr}^2}$, 
where c is the speed of light, $\lambda_0$ is the central wavelength, and $v_{\rm instr}$ is the instrumental line broadening. This varies with the wavelength,
 as the LETG resolution is $\sim$ 0.05 \AA. 
 The resulting parameters for the blue shift velocity,
 broadening and optical depth at which each line is formed
 are shown in Table~\ref{table:absorption}.
If our line identification is right, 
there is a large spread of blue-shift velocities, from 370 to 3100 km s$^{-1}$. While in some
other novae, such as V2491 Cyg, RS Oph \citep{2010AN....331..179N} and SMC 2016
\citep{Orio2018}, the blue-shift velocity
of the absorption lines has a narrow variation range, there are at least three velocity systems
 in the spectra of V4743 Sgr \citep{2011ApJ...733...70N}.

 The broadening of the lines also varies, and is larger than the instrumental
width and broadening (expected to be less than $\simeq$ 300 km s$^{-1}$),
 so like in V2491 Cyg \citep{2011ApJ...733...70N} the lines 
 may have been produced in a region of plasma
 with a significant extension, allowing us to view
a range of expansion velocities and thus a range of different
plasma layers.
\begin{table*}
\begin{flushleft}
\caption[Results from absorption line profiles measurements (KT Eri)]{Results of the fit to
 the absorption lines: measured wavelength, line shifts, widths, and optical depth. The errors are at a 90 percent confidence level.
}
\begin{minipage}{175mm}
\label{table:absorption}
\begin{tabular}{cccccc|cccc}
\hline
Ion & $\lambda_0$ & $\lambda_m$ & $v_{\rm shift}$ & $v_{\rm width}$ & $\tau_{\rm c}$ & $\lambda_m$ & $v_{\rm shift}$ & $v_{\rm width}$ & $\tau_{\rm c}$ \\
& (\AA) & (\AA) & (km\,s$^{-1}$) & (km\,s$^{-1}$) & & (\AA) & (km\,s$^{-1}$) & (km\,s$^{-1}$) \\
\hline
\multicolumn{6}{c|}{\bf Day\,71.3}&\multicolumn{4}{c}{\bf Day\,79.3}\\
\hline
N VII  & 24.779 & $...$ & $...$ &  $...$ &  $...$ & $24.520 \pm 0.014$ & \mbox{ $-3127 \pm 168$} & \mbox{ $1496 \pm 219$} & \mbox{ $0.54 \pm 0.05$} \\
N VI  & 24.898 & $...$ & $...$ &  $...$ &  $...$ & $24.718 \pm 0.014$ & \mbox{ $-2169 \pm 170$} & \mbox{ $1404 \pm 161$} & \mbox{ $0.54 \pm 0.05$} \\
Ca XII   & 27.973 & $27.849 \pm 0.003$ & $-1332 \pm 35$ &  $651 \pm 81$ &  $0.05 \pm 0.01$ & $27.898 \pm 0.008$ & \mbox{ $-801 \pm 87$} & \mbox{ $700 \pm 156$} & \mbox{ $0.19 \pm 0.03$} \\
Ar XV  & 28.346 & $28.219 \pm 0.002$ & $-1341 \pm 198$ &  $431 \pm 41$ &  $0.04 \pm 0.01$ & $28.268 \pm 0.005$ & \mbox{ $-828 \pm 48$} & \mbox{ $556 \pm 82$} & \mbox{ $0.20 \pm 0.02$} \\
Ca XII    & 28.681 & $...$ & $...$ &  $...$ &  $...$ & $28.510 \pm 0.005$ & \mbox{ $-1784 \pm 52$} & \mbox{ $834 \pm 97$} & \mbox{ $0.42 \pm 0.03$} \\
N VI r  & 28.787 & $...$ & $...$ &  $...$ &  $...$ & $28.706 \pm 0.003$ & \mbox{ $-843 \pm 33$} & \mbox{ $401 \pm 54$} & \mbox{ $0.25 \pm 0.03$} \\
Ar XIII   & 29.497 & $29.190 \pm 0.004$ & $-3119 \pm 45$ &  $456 \pm 95$ &  $0.08 \pm 0.01$ & $29.239 \pm 0.006$ & \mbox{ $-2627 \pm 59$} & \mbox{ $1100 \pm 115$} & \mbox{ $0.52 \pm 0.04$} \\
N VI  & 29.535 & $29.467 \pm 0.003$ & $-687 \pm 32$ &  $817 \pm 63$ &  $0.08 \pm 0.01$ & $29.499 \pm 0.006$ & \mbox{ $-367 \pm 62$} & \mbox{ $585 \pm 111$} & \mbox{ $0.22 \pm 0.03$} \\
Ca XI   & 30.448 & $30.348 \pm 0.005$ & $-981 \pm 46$ &  $1053 \pm 99$ &  $0.09 \pm 0.01$ & $30.356 \pm 0.004$ & \mbox{ $-909 \pm 41$} & \mbox{ $641 \pm 68$} & \mbox{ $0.33 \pm 0.03$} \\
C V   & 32.400 & $32.100 \pm 0.004$ & $-2779 \pm 40$ &  $787 \pm 81$ &  $0.07 \pm 0.01$ & $32.147 \pm 0.005$ & \mbox{ $-2338 \pm 45$} & \mbox{ $543 \pm 68$} & \mbox{ $0.32 \pm 0.03$} \\
C VI  & 33.734 & $33.618 \pm 0.004$ & $-1027 \pm 38$ &  $662 \pm 78$ &  $0.09 \pm 0.01$ & $33.661 \pm 0.007$ & \mbox{ $-644 \pm 66$} & \mbox{ $600 \pm 130$} & \mbox{ $0.15 \pm 0.02$} \\

\hline
\multicolumn{6}{c|}{\bf Day\,84.6}&\multicolumn{4}{c}{\bf Day\,158.8}\\
\hline
N VII  & 24.779 & $...$ & $...$ &  $...$ &  $...$ & $24.544 \pm 0.012$ & \mbox{ $-2846 \pm 141$} & \mbox{ $1188 \pm 120$} & \mbox{ $0.44 \pm 0.05$} \\
N VI  & 24.898 & $...$ & $...$ &  $...$ &  $...$ & $24.706 \pm 0.007$ & \mbox{ $-2315 \pm 88$} & \mbox{ $1321 \pm 96$} & \mbox{ $0.79 \pm 0.04$} \\
Ca XII   & 27.973 & $27.866 \pm 0.006$ & $-1144 \pm 70$ &  $679 \pm 138$ &  $0.16 \pm 0.02$ & $27.895 \pm 0.013$ & \mbox{ $-835 \pm 144$} & \mbox{ $809 \pm 347$} & \mbox{ $0.23 \pm 0.10$} \\
Ar XV  & 28.346 & $28.233 \pm 0.002$ & $-1190 \pm 26$ &  $423 \pm 41$ &  $0.18 \pm 0.01$ & $28.238 \pm 0.004$ & \mbox{ $-1142 \pm 44$} & \mbox{ $432 \pm 73$} & \mbox{ $0.22 \pm 0.03$} \\
Ca XII   & 28.681 & $28.529 \pm 0.003$ & $-1589 \pm 32$ &  $308 \pm 50$ &  $0.17 \pm 0.02$ & $28.545 \pm 0.005$ & \mbox{ $-1422 \pm 56$} & \mbox{ $376 \pm 99$} & \mbox{ $0.18 \pm 0.03$} \\
N VI r  & 28.787 & $28.701 \pm 0.004$ & $-892 \pm 40$ &  $803 \pm 78$ &  $0.41 \pm 0.03$ & $28.693 \pm 0.003$ & \mbox{ $-977 \pm 32$} & \mbox{ $474 \pm 51$} & \mbox{ $0.38 \pm 0.03$} \\
Ar XIII   & 29.497 & $29.213 \pm 0.003$ & $-2885 \pm 26$ &  $791 \pm 44$ &  $0.49 \pm 0.02$ & $29.227 \pm 0.006$ & \mbox{ $-2744 \pm 59$} & \mbox{ $1078 \pm 120$} & \mbox{ $0.46 \pm 0.04$} \\
N VI  & 29.535 & $29.460 \pm 0.003$ & $-764 \pm 35$ &  $707 \pm 65$ &  $0.33 \pm 0.02$ & $29.458 \pm 0.007$ & \mbox{ $-780 \pm 72$} & \mbox{ $498 \pm 125$} & \mbox{ $0.24 \pm 0.04$} \\
Ca XI   & 30.448 & $30.343 \pm 0.004$ & $-1036 \pm 39 $ &  $788 \pm 70$ &  $0.33 \pm 0.02$ & $30.330 \pm 0.003$ & \mbox{ $-1160 \pm 27$} & \mbox{ $595 \pm 44$} & \mbox{ $0.49 \pm 0.03$} \\
C V   & 32.400 & $32.098 \pm 0.004$ & $-2795 \pm 34$ &  $679 \pm 53$ &  $0.25 \pm 0.02$ & $32.122 \pm 0.005$ & \mbox{ $-2569 \pm 44$} & \mbox{ $551 \pm 66$} & \mbox{ $0.35 \pm 0.03$} \\
C VI  & 33.734 & $33.632 \pm 0.006$ & $-911 \pm 52$ &  $533 \pm 86$ &  $0.16 \pm 0.02$ & $33.614 \pm 0.006$ & \mbox{ $-1070 \pm 56$} & \mbox{ $565 \pm 86$} & \mbox{ $0.17 \pm 0.02$} \\
\hline
\end{tabular}
\end{minipage}
\end{flushleft}
\end{table*}
\begin{figure*}
\begin{center}
\includegraphics[width=8.5cm]{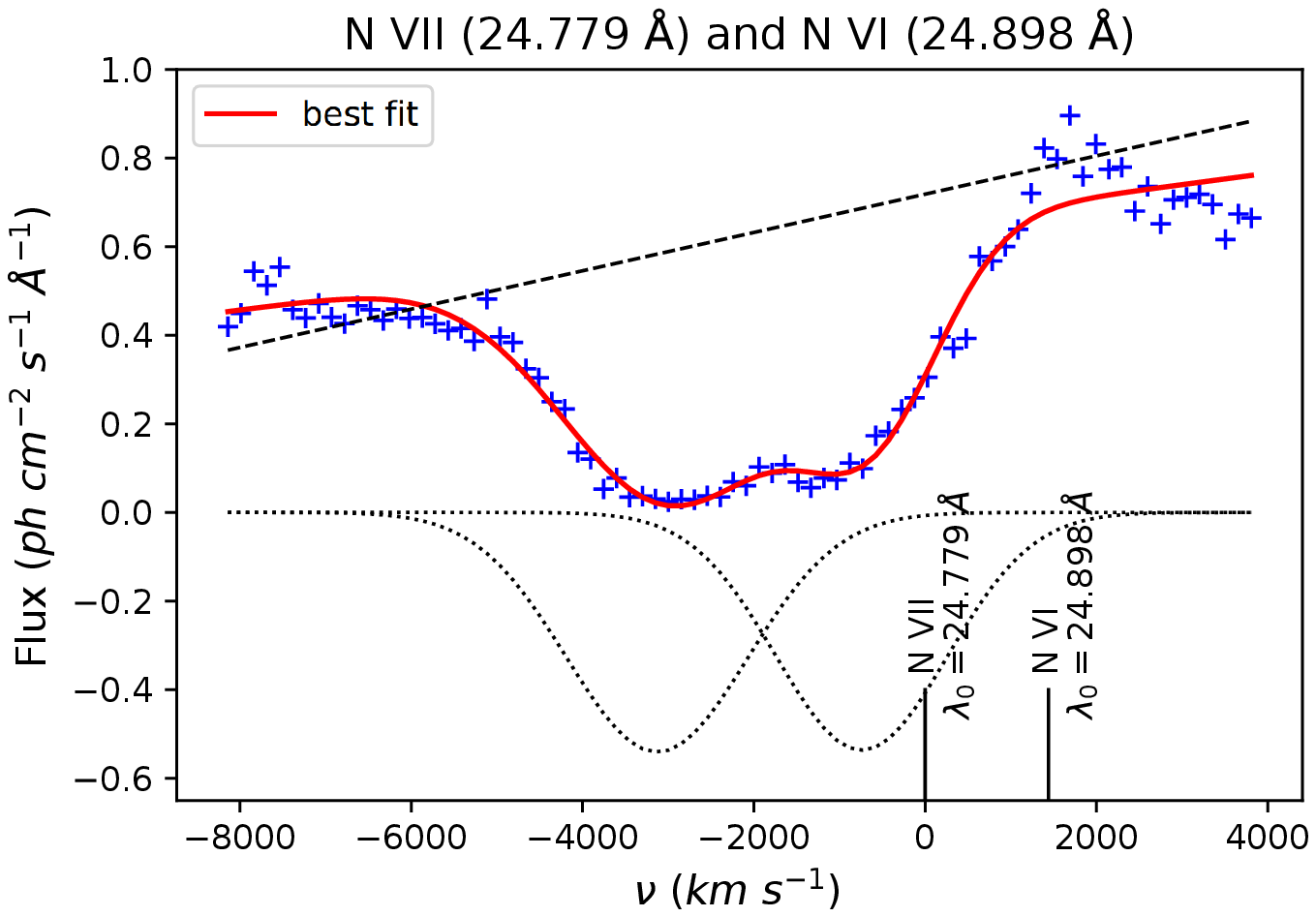}
\hspace{0.08em}
\includegraphics[width=8.5cm]{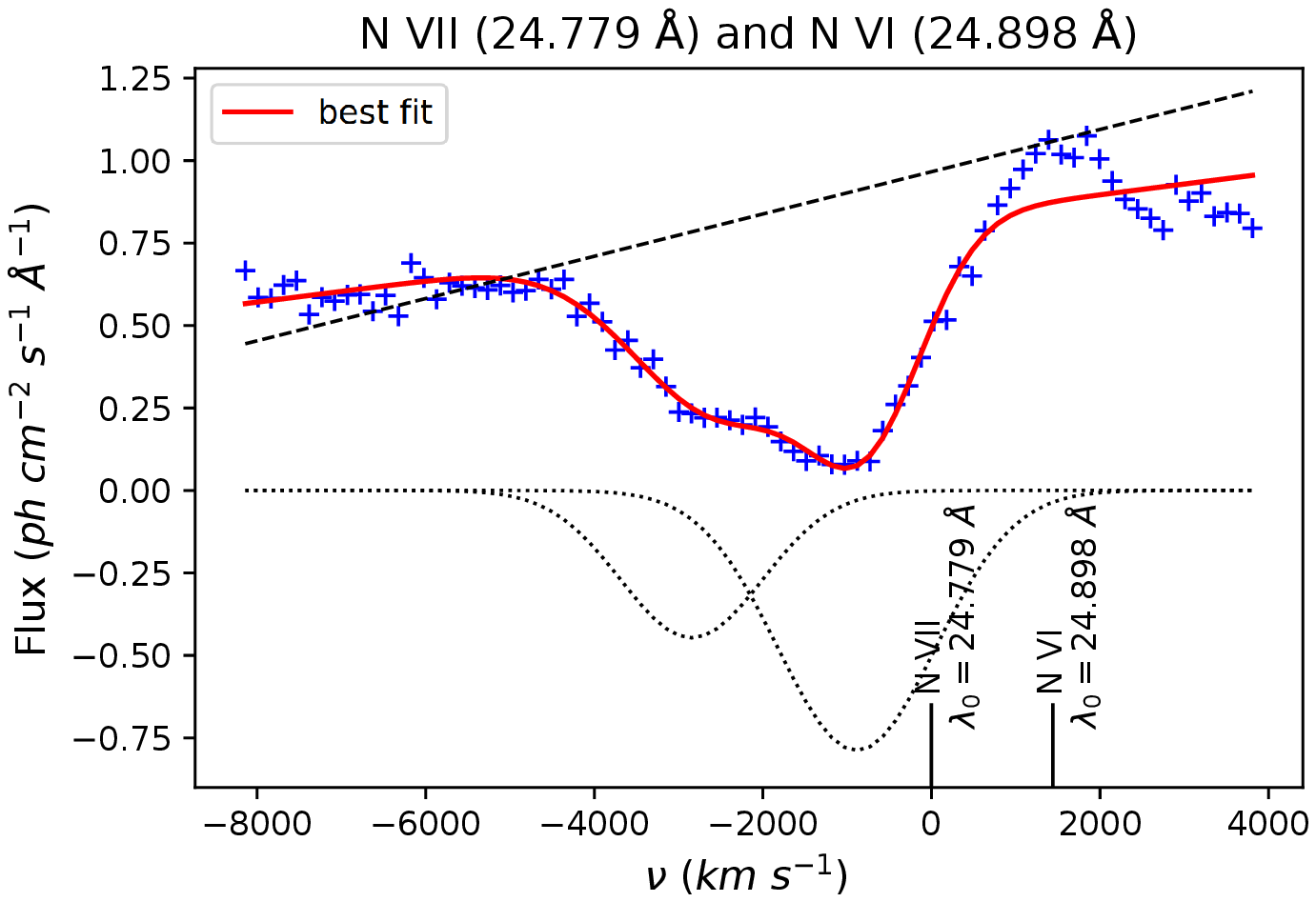}
\caption{Fit to the blended lines of N VII ($\lambda_0=24.779$\,\AA) and N VI ($\lambda_0=24.898$\,\AA) for the spectra of KT Eridani on days 79.3 (left) and 158.8 (right). The blue crosses represent the observed flux and its error bar,
 the solid red lines represent the best-fit by using the method described by \citet{2010AN....331..179N, 2011ApJ...733...70N}, and the dotted black lines and dashed black line represent the Gaussian-like component and continuum component in the spectra, respectively.}\label{fig: line fits}
\end{center}
\end{figure*}
\begin{figure*}
\begin{center}
\includegraphics[width=5.6cm]{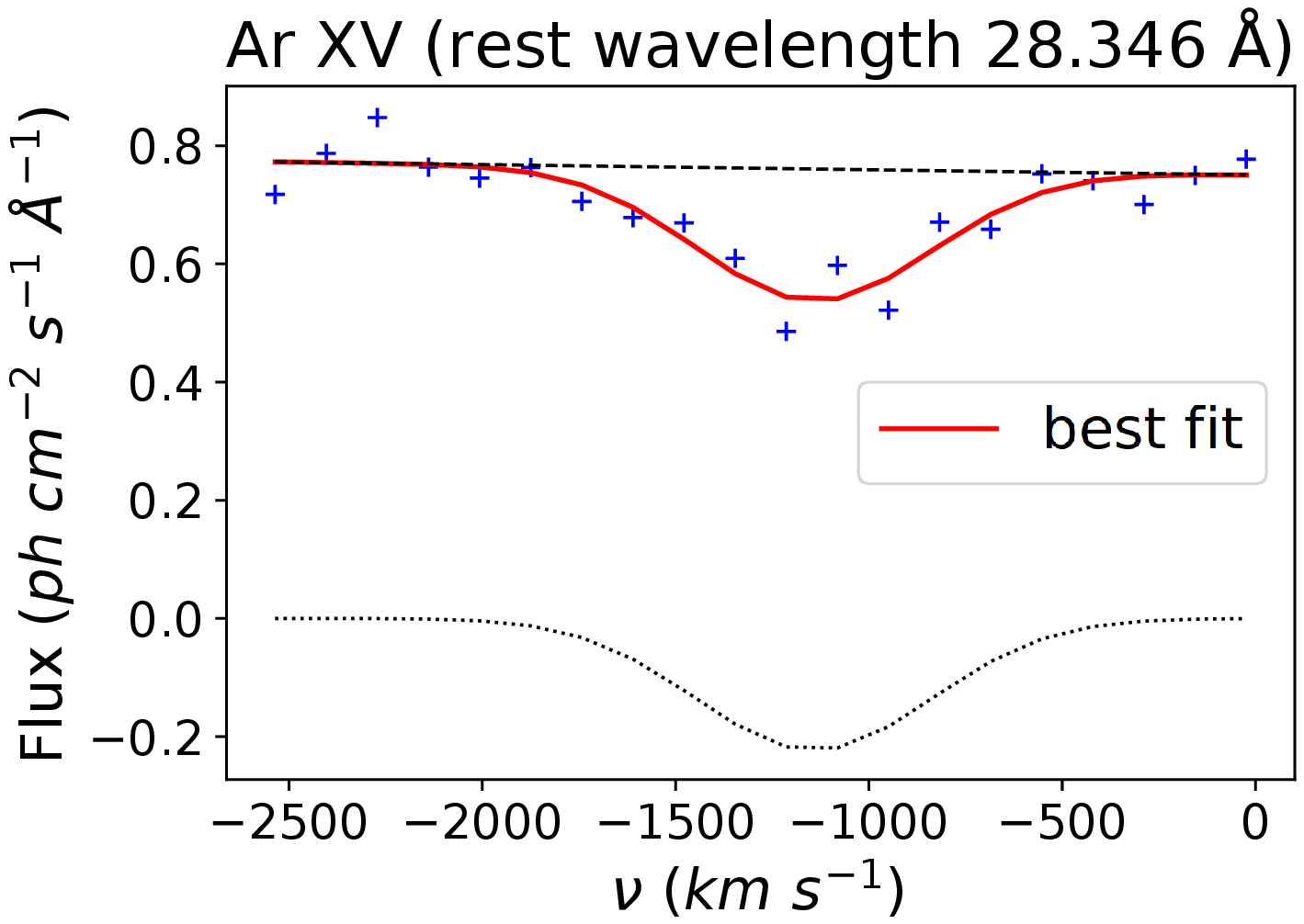}
\hspace{-0.01em}
\includegraphics[width=5.6cm]{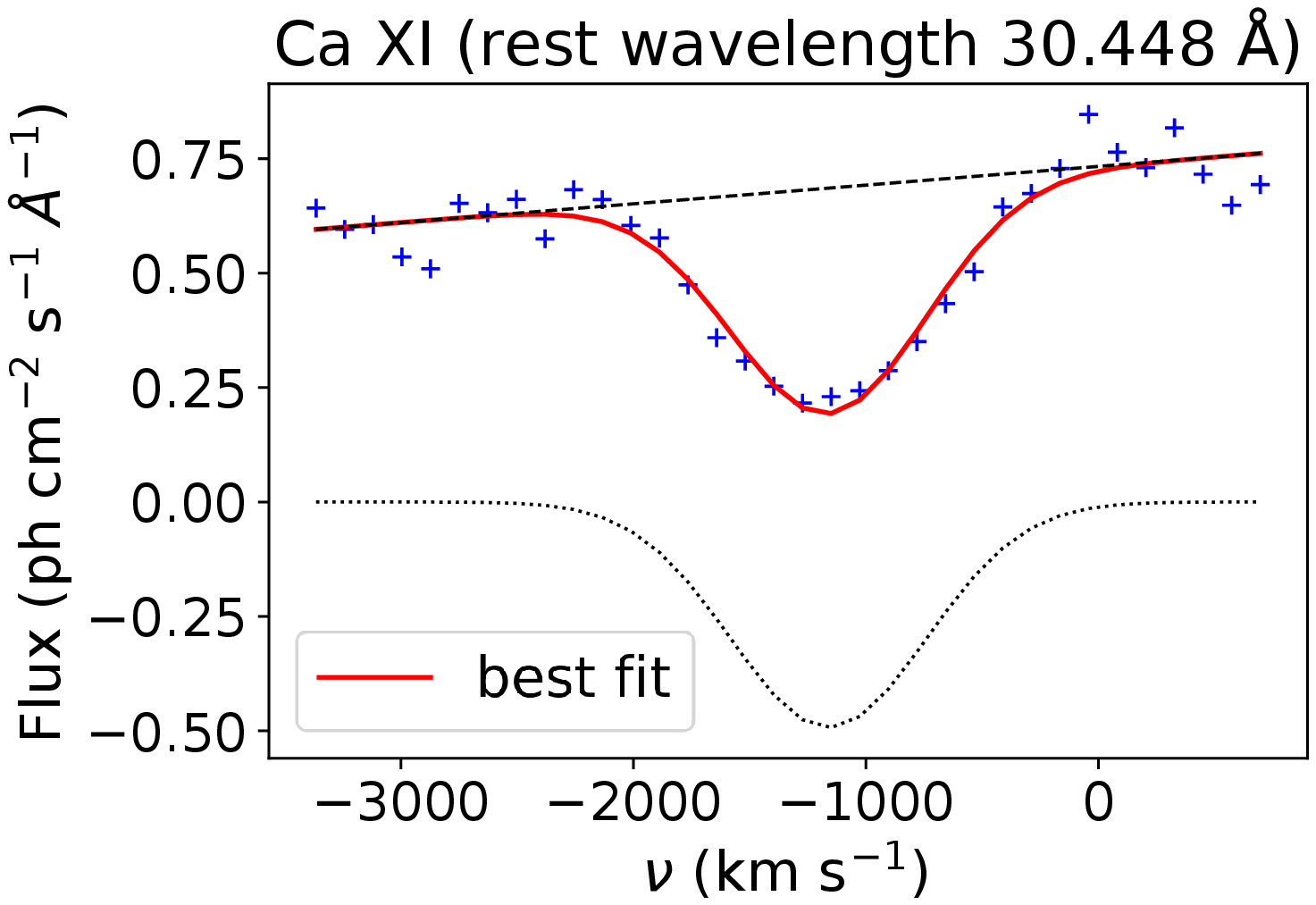}
\hspace{-0.01em}
\includegraphics[width=5.6cm]{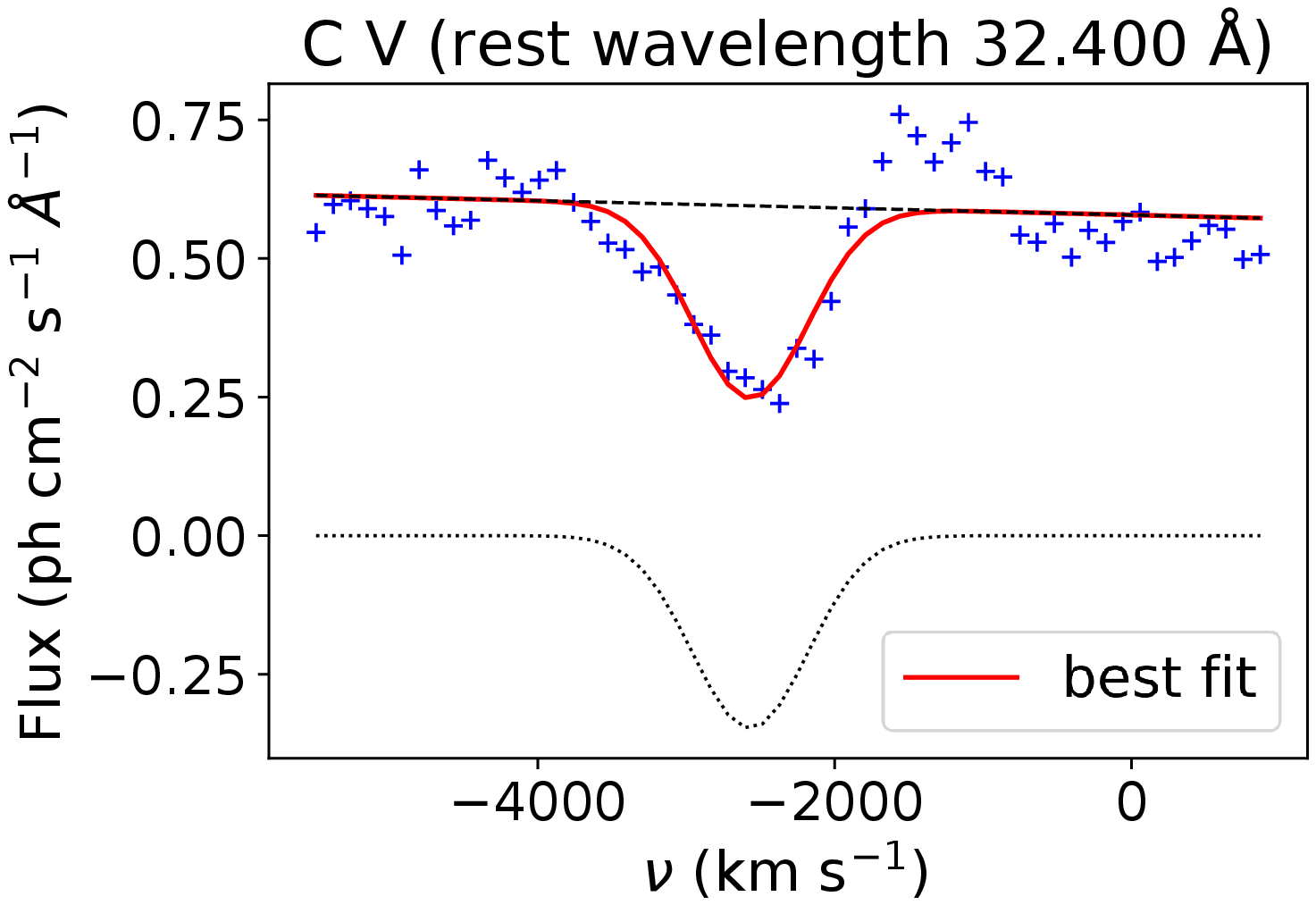}
\caption{Fit to the absorption lines of Ar XV ($\lambda_0=28.346$\,\AA), Ca XI ($\lambda_0=30.448$\,\AA), and C V ($\lambda_0=32.400$\,\AA) on day 158.8. The blue cross symbols represent the observed spectra with error bar, the solid red lines represent the best-fit by using the method described by \citet{2010AN....331..179N, 2011ApJ...733...70N}, and the dotted black lines and dashed black line represent the Gaussian-like component and continuum component in the spectra, respectively.}\label{fig: line fits 2}
\end{center}
\end{figure*}
The
 absorption lines of N VII ($\lambda_0=24.779$\,\AA) and N VI ($\lambda_0=24.898$\,\AA) are blended,
 generating a flat bottom in the spectrum of day 84.6.
Moreover, the N VI ($\lambda_0=24.898$\,\AA) has the largest optical depth it is saturated in all spectra. 
We derived wavelength, line shifts, widths, and optical depths by
 assuming an overlap of the lines, as shown in Fig.~\ref{fig: line fits}, and
 propose the possible fit for the spectra of days 79.3 and 158.8, but in the third observation
 the lines are detected, yet we were not able to obtain a fit.
Even if the fit is acceptable, this blending introduces a large uncertainty.

We also identified emission lines of nitrogen, carbon and sulfur in all four spectra. As shown in Table~\ref{table:emission}, these emission lines are red-shifted with velocities in a large range, 1700-3200 km s$^{-1}$.
We used a Gaussian model (AGAUSS in XSPEC) to fit the profiles of the emission lines
 to determine the measured wavelength, line width and line shifts, modeling the 
underlying continuum with TMAP as shown in the next Section (even in this case, however, the
 choice of continuum is not critical).
 The AGAUSS model in XSPEC computes the line profile as follows:
\begin{equation}
\label{eq:eq2}
A(\lambda)=\frac{K}{\sigma \sqrt{2 \pi}}exp{(-(\lambda-\lambda_{l})^{2}/2 \sigma^{2})}
\end{equation}
where $\lambda_{l}$ and $\sigma$ are the line central wavelength in Angstrom and the instrumental broadening corrected line width in Angstrom, respectively.
The resulting parameters are shown in Table~\ref{table:emission}. Table~\ref{table:highlow} also shows the difference in flux for the ``high'' and ``low'' periods defined for Fig.~\ref{fig: timecompspecfit2} for day 71.3. The emission lines that are strong enough to be identified are in the wavelength band $\sim$ 24 $-$ 37 \AA.
The emission line N VII ($\lambda_0=24.779$\,\AA) merge in the continuum emission and is too weak to be identified in spectrum of day 71.3 which has the lowest count rate, but it can be identified in the spectra of the rest three observations. It is weak on day 84.6 when the count rate is low, and it is strong on days 79.3 and 158.8 when the count rates are high. The flux of emission line N VI r ($\lambda_0=28.787$\,\AA) on day 84.6 is higher than day 158.8 when the count rate is the highest. Therefore, the fluxes of the emission lines do not change monotonically with the count rate of the spectra.

\citet{2013A&A...559A..50N} suggest that the emission lines in
 V2491 Cyg originated from the reprocessed emission far
 from the WD, in the clumpy ejecta of
 the nova wind. We also attribute them to the ejecta in KT Eri. 
\begin{table*}
\begin{flushleft}
\caption[Parameters of the fit to the emission lines]{Parameters
 of the Gaussian fit
 to the emission lines:
measured wavelength, line shifts, widths, and fluxes $\times$ $10^{-11}$erg cm$^{-2}$ s$^{-1}$. 
 The fluxes are calculated with the cflux command in XSPEC. The errors are at a 90 percent confidence level.}
\begin{minipage}{175mm}
\label{table:emission}
\begin{tabular}{cccccc|cccc}
\hline
Ion & $\lambda_0$ & $\lambda_m$   & $v_{\rm shift}$ & $v_{\rm width}$ & Flux & $\lambda_m$ & $v_{\rm shift}$  & $v_{\rm width}$ & Flux \\
& (\AA) & (\AA) & (km\,s$^{-1}$) & (km\,s$^{-1}$)  & & (\AA) & (km\,s$^{-1}$) & (km\,s$^{-1}$)  \\
\hline
\multicolumn{6}{c|}{\bf Day\,71.3}&\multicolumn{4}{c}{\bf Day\,79.3}\\
\hline
N VII  & 24.779 & ... & ... &  ... & ... &  24.923$^{+0.006}_{-0.005}$  & 1747$^{+74}_{-64}$ & 1368$^{+137}_{-119}$   & 9.55$^{+0.54}_{-0.76}$  \\
C VI  & 25.830 & 26.087$^{+0.004}_{-0.003}$  & 2987$^{+43}_{-39}$ & 640$^{+65}_{-50}$  &  0.80$^{+0.05}_{-0.05}$ &  26.079$^{+0.008}_{-0.019}$  & 2884$^{+95}_{-223}$  & 705$^{+214}_{-164}$  & 1.85$^{+0.33}_{-0.33}$  \\
S XIII  & 27.392 & 27.653$^{+0.005}_{-0.003}$  & 2856$^{+54}_{-30}$ & 882$^{+46}_{-47}$  &  1.00$^{+0.14}_{-0.07}$ &  27.675$^{+0.007}_{-0.006}$   & 3016$^{+77}_{-63}$  & 433$^{+93}_{-78}$    & 2.65$^{+0.38}_{-0.39}$  \\
N VI r   & 28.787 & 29.054$^{+0.004}_{-0.004}$    & 2780$^{+44}_{-41}$ & 1281$^{+44}_{-59}$  &  1.18$^{+0.09}_{-0.08}$ &  29.047$^{+0.005}_{-0.004}$  & 2783$^{+54}_{-39}$ & 839$^{+59}_{-44}$    & 8.93$^{+0.54}_{-0.54}$  \\
N VI & 29.084 & 29.348  & 2721   & 583    &  0.16$^{+0.06}_{-0.06}$ &  29.395$^{+0.003}_{-0.009}$  & 3205$^{+34}_{-92}$  & 364$^{+73}_{-73}$  &  2.92$^{+0.33}_{-0.32}$  \\
S XII  & 36.398 & 36.767$^{+0.005}_{-0.007}$   & 3039$^{+42}_{-67}$  & 673$^{+91}_{-76}$   &  0.65$^{+0.06}_{-0.06}$ &  36.756$^{+0.017}_{-0.016}$   & 2948$^{+140}_{-132}$   & 509$^{+294}_{-230}$   &  0.62$^{+0.23}_{-0.20}$  \\
\hline
\multicolumn{6}{c|}{\bf Day\,84.6}&\multicolumn{4}{c}{\bf Day\,158.8}\\
\hline
N VII  & 24.779  & 24.950$^{+0.008}_{-0.008}$   & 2069$^{+102}_{-96}$ & 328$^{+151}_{-209}$   &  0.89$^{+0.19}_{-0.18}$ &  24.947$^{+0.003}_{-0.010}$   & 2040$^{+35}_{-127}$ & 1914$^{+96}_{-96}$   & 16.40$^{+0.61}_{-0.60}$  \\
C VI  & 25.830 & 26.056$^{+0.005}_{-0.008}$   & 2624$^{+55}_{-95}$  & 525$^{+115}_{-99}$   &  2.10$^{+0.25}_{-0.24}$ &  26.038$^{+0.013}_{-0.012}$  & 2416$^{+151}_{-144}$  & 574$^{+181}_{-197}$  & 1.57$^{+0.40}_{-0.39}$  \\
S XIII  & 27.392 & 27.680$^{+0.007}_{-0.004}$   & 3147$^{+72}_{-44}$  & 1021$^{+77}_{-62}$   &  6.39$^{+0.50}_{-0.50}$ &  27.638$^{+0.051}_{-0.072}$   & 2696$^{+567}_{-790}$  & 1801$^{+1127}_{-526}$   & 7.69$^{+1.25}_{-1.25}$  \\
N VI r  & 28.787 & 29.031$^{+0.004}_{-0.003}$  & 2610$^{+129}_{-129}$  & 1075$^{+44}_{-59}$    &  9.42$^{+0.42}_{-0.41}$ &  29.030$^{+0.013}_{-0.012}$   & 2531$^{+143}_{-127}$  & 1016$^{+162}_{-148}$  & 3.99$^{+0.67}_{-0.66}$  \\
N VI & 29.084 & 29.346$^{+0.005}_{-0.005}$   & 2708$^{+53}_{-54}$  & 192$^{+102}_{-102}$   &  1.78$^{+0.26}_{-0.26}$ &  29.372$^{+0.004}_{-0.006}$   & 2969$^{+41}_{-63}$   & 244$^{+72}_{-73}$   &  2.78$^{+0.35}_{-0.34}$  \\
S XII  & 36.398 & 36.786$^{+0.013}_{-0.012}$   & 3195$^{+108}_{-99}$  & 656$^{+179}_{-147}$   &  0.96$^{+0.22}_{-0.20}$ &  36.730   & 2734   & 467   &  0.20$^{+0.39}_{-0.09}$  \\
\hline
\end{tabular}
\end{minipage}
\end{flushleft}
\end{table*}
\begin{table}
\caption[Flux in the ``high'' and ``low'' periods]{Flux resulting from
 the Gaussian fit to the emission lines, in units of 
$10^{-11}$erg cm$^{-2}$ s$^{-1}$, for the periods of higher than average
 count rate and for the periods of lower than average
 count rate on day 71.3, calculated with the cflux command in XSPEC. The errors are at a 90 percent confidence level.}
\begin{minipage}{175mm}
\label{table:highlow}
\begin{tabular}{cccc}
\hline
Ion & $\lambda_0$ &  Flux (high) & Flux (low) \\
& (\AA) &   & \\
\hline
\multicolumn{4}{c|}{\bf Day\,71.3}\\
\hline
N VII  & 24.779  & ... &  ...    \\
C VI  & 25.830  & 0.86$^{+0.13}_{-0.14}$ & 0.39$^{+0.09}_{-0.09}$  \\
S XIII  & 27.392  & 1.48$^{+0.14}_{-0.14}$ & 0.82$^{+0.12}_{-0.12}$ \\
N VI r   & 28.787     & 1.56$^{+0.15}_{-0.16}$ & 0.85$^{+0.12}_{-0.13}$  \\
N VI & 29.084  &   0.28$^{+0.11}_{-0.11}$ & ...  \\
S XII   & 36.398     & 0.91$^{+0.14}_{-0.15}$ & 0.49$^{+0.08}_{-0.09}$  \\
\hline
\end{tabular}
\end{minipage}
\end{table}
%
\section{Fits with comprehensive physical models}
\subsection{Fit with the TMAP library of atmospheric models} \label{sec:fit}
\begin{figure*}
\begin{center}
\includegraphics[width=8.2cm,clip]{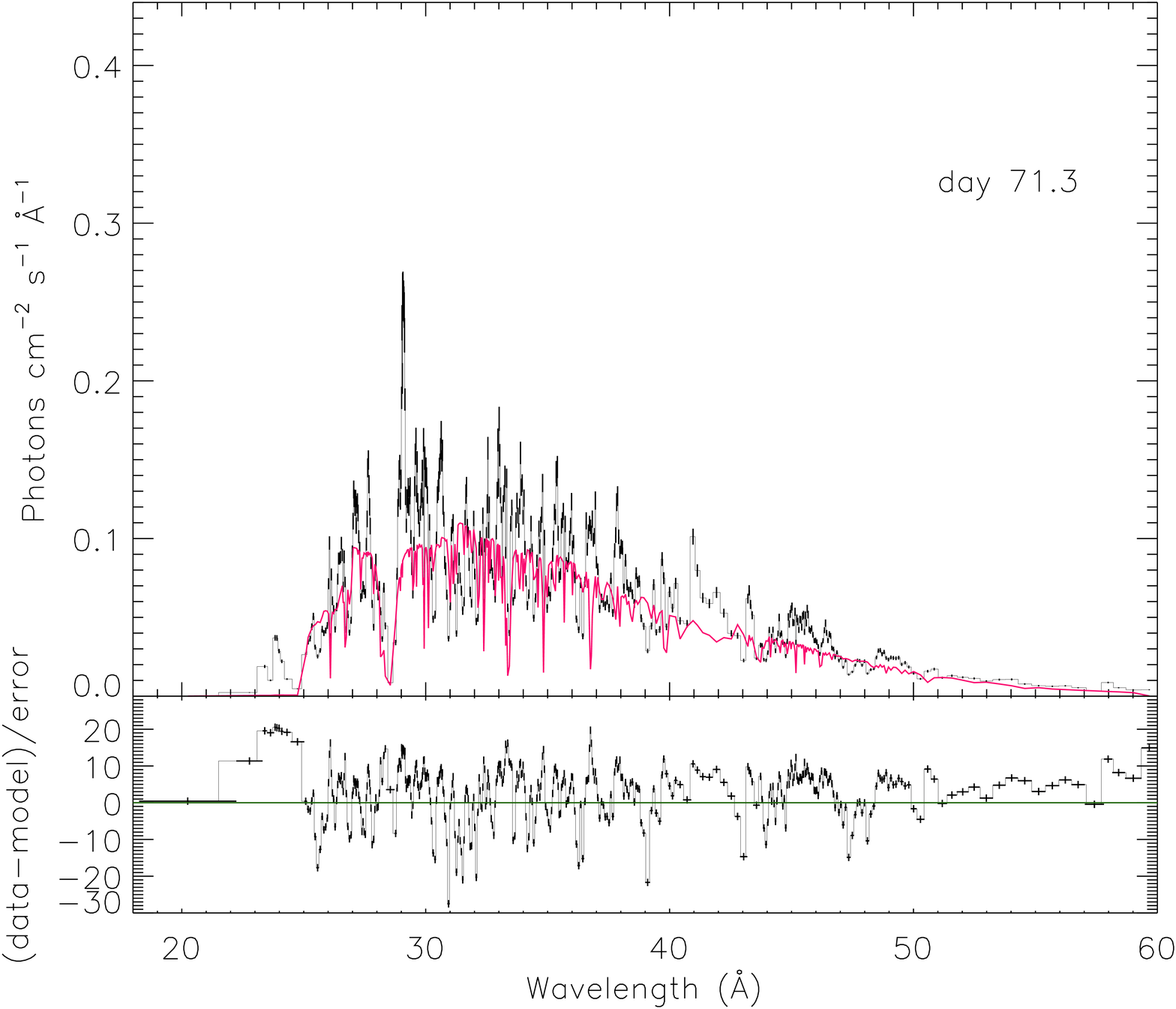}
\hspace{0.22cm}
\includegraphics[width=8.2cm,clip]{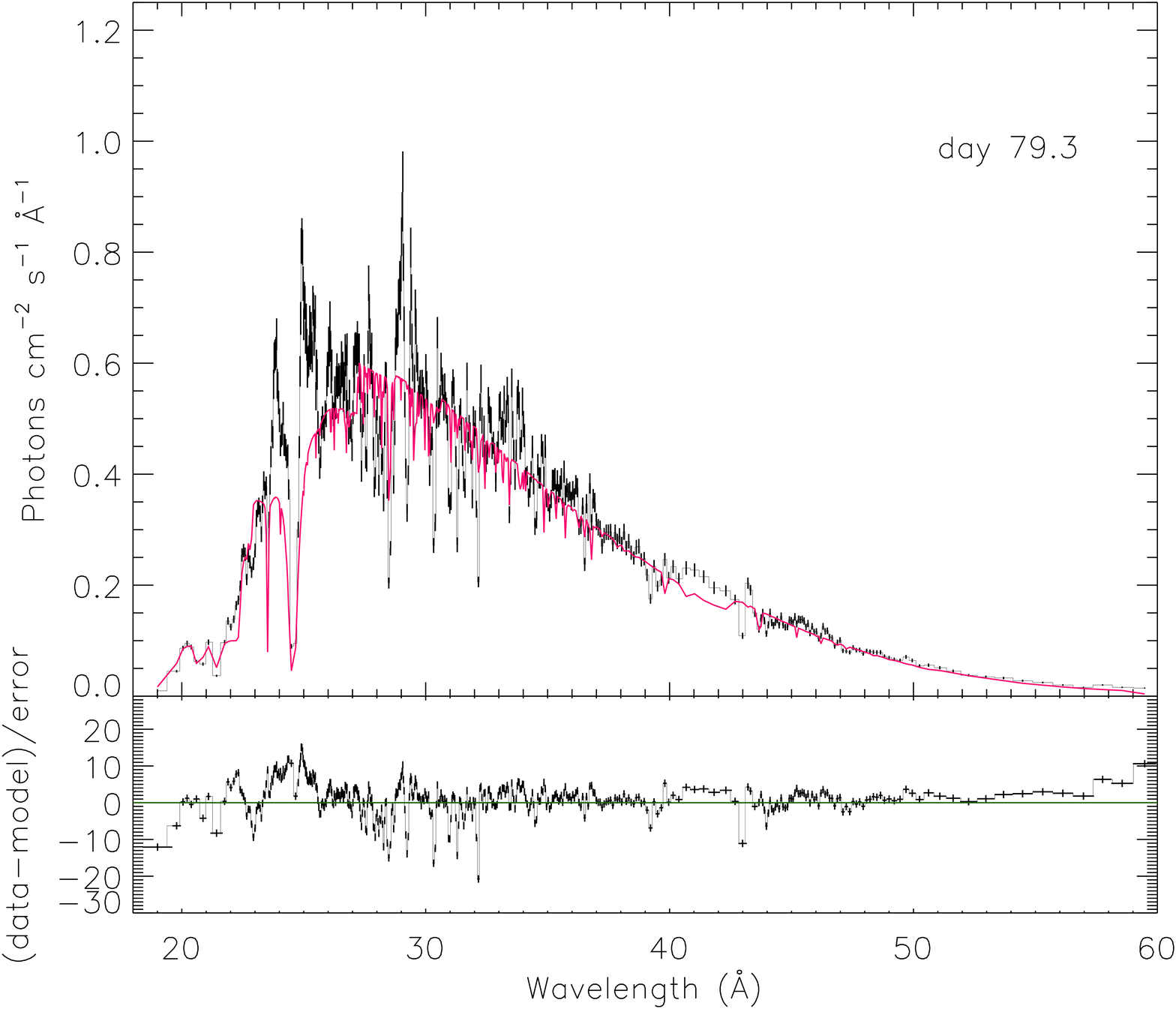}
\includegraphics[width=8.2cm,clip]{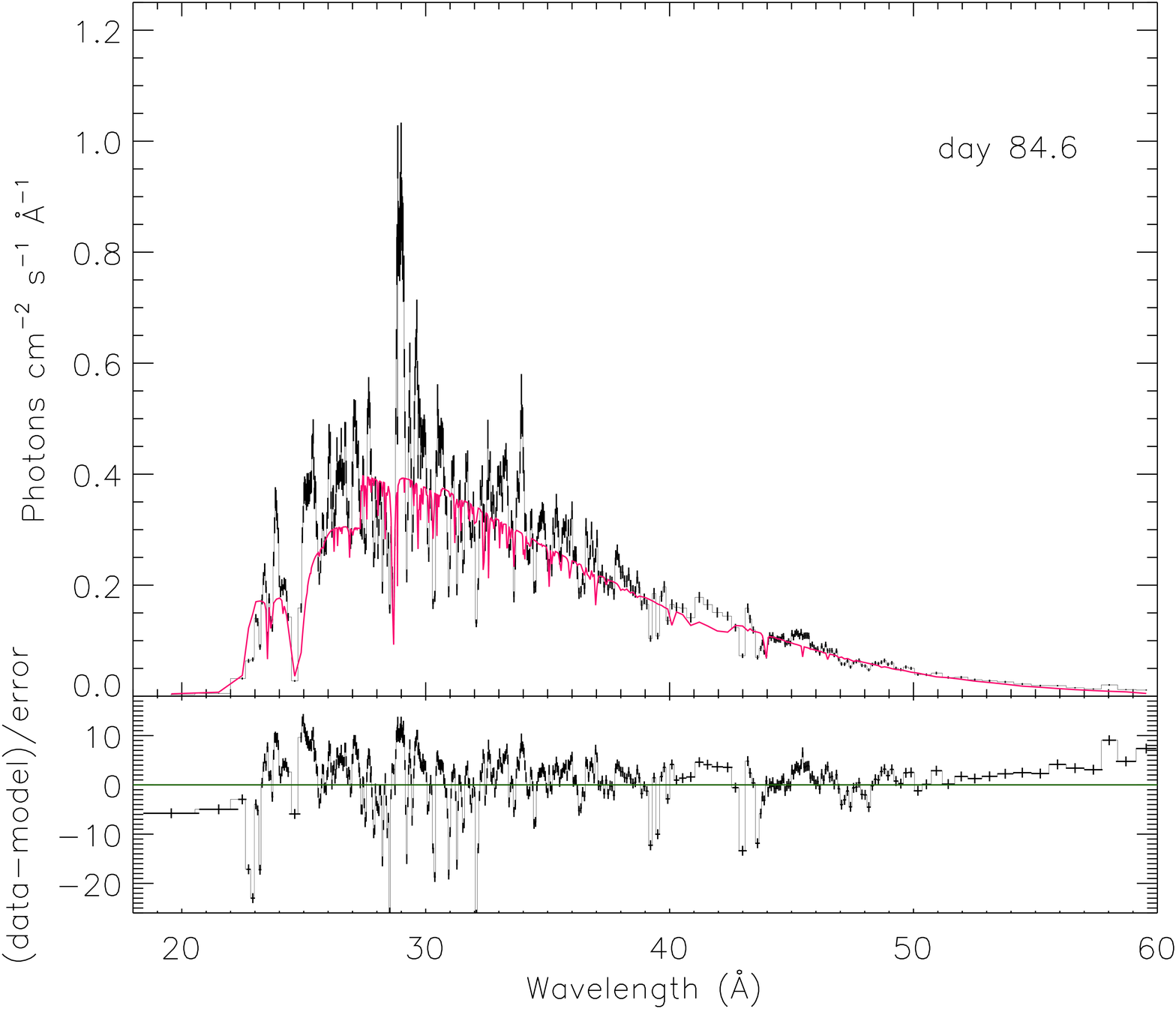}
\hspace{0.22cm}
\includegraphics[width=8.2cm,clip]{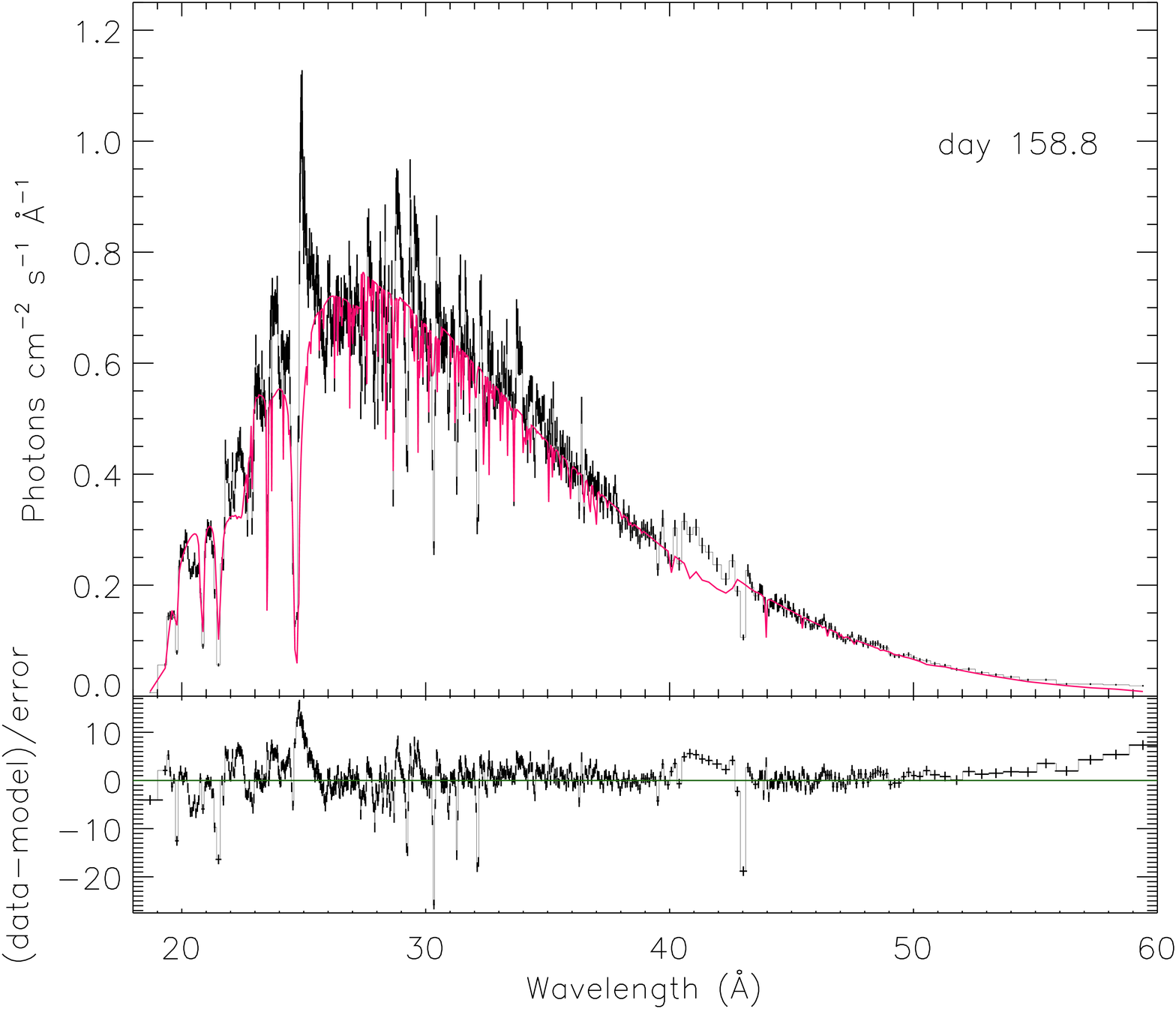}
\caption{The upper panels show
 the TBABS+TMAP fits to the spectra of the four epochs, traced in solid red lines. 
 The corresponding residuals are shown in each lower panel. 
 For the spectrum of day 71.3, 
N(H) is fixed at 6.00 $\times$ 10$^{20}$ cm$^{-2}$. log(g)=9
 in all fits. All the other
parameters are reported in Table~\ref{table:parameters TMAP}.
 The abundances in the table correspond to models: 201 (SSS$_{-}$201$_{-}$00010-00060.bin$_{-}$0.002$_{-}$9.00.fits) for day 71.3, 003 (SSS$_{-}$003$_{-}$00010-00060.bin$_{-}$0.002$_{-}$9.00.fits) for
 days 79.3 and 84.6, and 004 (SSS$_{-}$004$_{-}$00010-00060.bin$_{-}$0.002$_{-}$9.00.fits) for day 158.8. The absorption edges tend
 to be overestimated. }\label{fig:timecompspecfit4}
\end{center}
\end{figure*}

\begin{figure*}
\begin{center}
\includegraphics[width=8.2cm,clip]{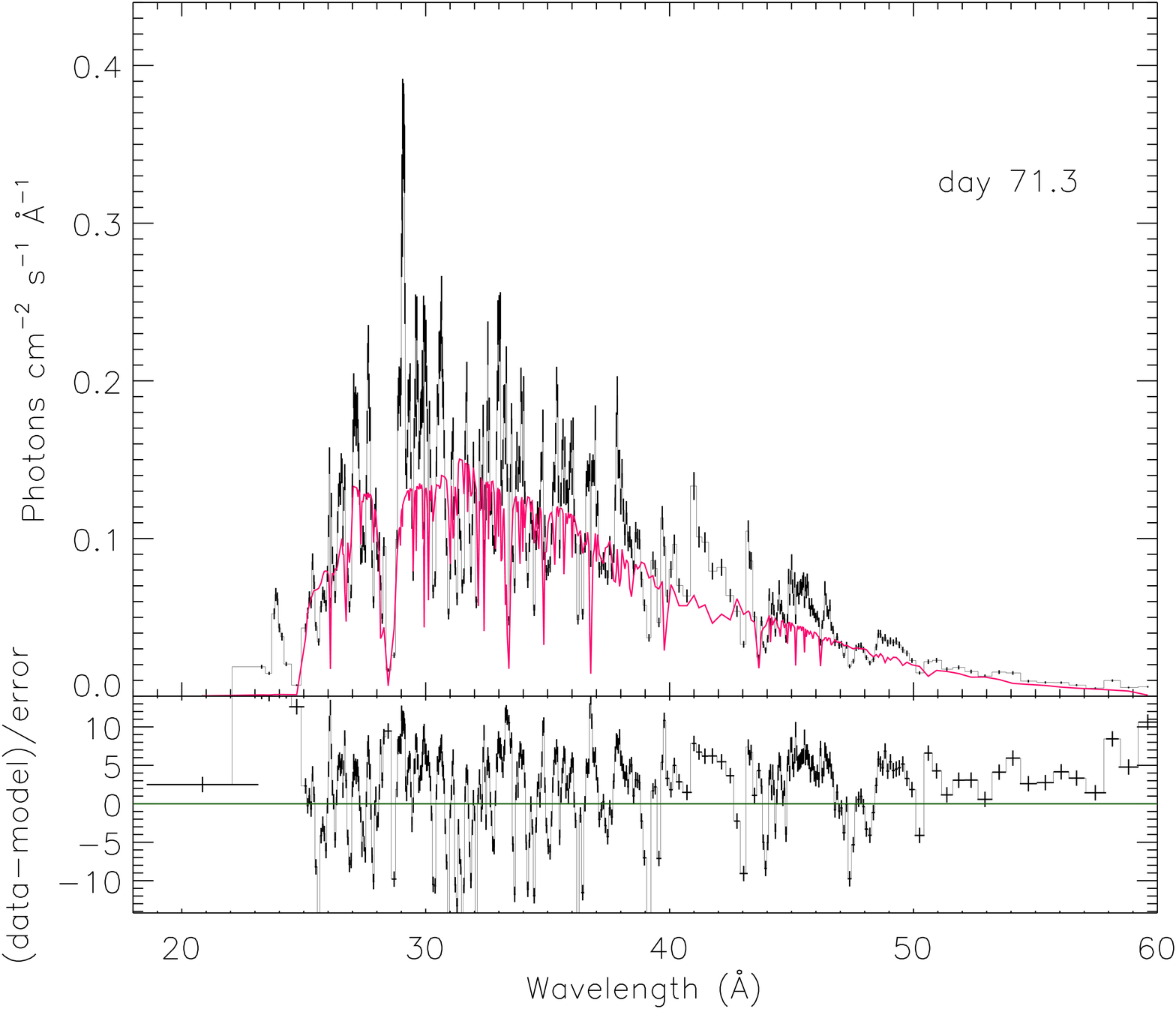}
\hspace{0.22cm}
\includegraphics[width=8.2cm,clip]{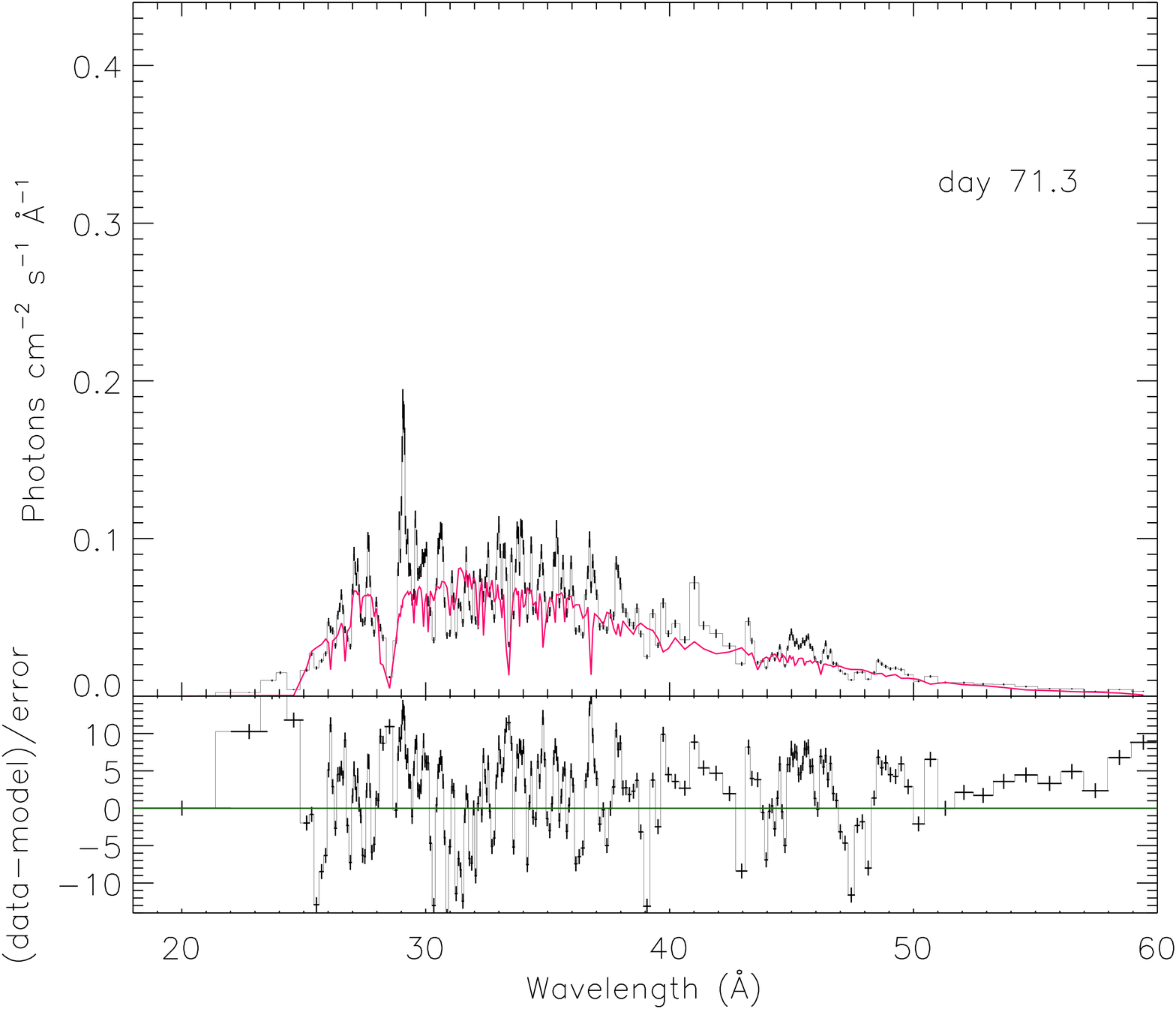}
\caption{Spectra of day 71.3 after the optical-maximum, fitted using TBABS 
 and TMAP (201) with log(g)=9. 
The corresponding residuals of each fit are shown in each lower panel.
On the left
 is the spectrum extracted when the zero order light curve count rate was larger than the average 13.87 counts s$^{-1}$.
The right panel shows the spectrum extracted when the zero order light curve count rate was lower than the average 13.87 counts s$^{-1}$. The parameters are reported in Table~\ref{table:parameters TMAP}.}\label{fig: timecompspecfit2}
\end{center}
\end{figure*}

 Fig.~\ref{fig:timecompspecfit4} shows the fit to the spectra of each exposure obtained with the fitting package
 XSPEC v 12.10.1 \citep{1996ASPC..101...17A}, with the TMAP model non-local thermodynamic equilibrium (NLTE) models \citep{2003A&A...403..709R, 2010ApJ...717..363R} 
for an extremely hot WD atmosphere, as expected above a shell of ongoing
 thermonuclear burning. We used a public grid of calculated models with effective gravity 
 indicated as log(g), varying between log(g)=5 and
log(g)=9 (with a step increment of 1).
 These models were included as tables in XSPEC, using the tabular
 additive model ATABLE. Because the absorption features are
 blue-shifted, we assumed a constant blue shift for each absorption features except those
 originating in the cool gas of the intervening ISM.
 We found that the nova is so hot that only models with log(g)=9 are
 relevant, or else the luminosity would be largely above Eddington value. 
The Tuebingen-Boulder ISM absorption model 
TBABS \citep{2000ApJ...542..914W} in XSPEC was used to take 
the hydrogen column into account.
Fig.~\ref{fig: timecompspecfit2} shows the fit to the ``higher than average'' and ``lower than average''
 spectra of the first observation.

We first note that, on the blue or ``hard'' side, the flux in
 the spectrum of the first exposure appears to be cut by
a K absorption edge of N VI (22.46 \AA) while for the other 3 observations, the 
 absorption edge that cuts the flux appears to be due to N VII at 18.59 \AA,
 and this indicates lower temperature in the first observations. It also constrains the temperature in the last three exposures to be higher, and within a close range.
The fits with TBABS+TMAP that minimize
 the $\chi^2$ parameters for all the observations are shown in Fig.~\ref{fig:timecompspecfit4} 
and Fig.~\ref{fig: timecompspecfit2},
 and the parameters are given in the Table~\ref{table:parameters TMAP}. 
The minimum $\chi^2$ parameter divided by degrees of freedom was much larger than 1
(around 3 for the last three observations), 
so we could not conclude that these ``best fits'' are statistically good. In fact, many 
 spectral features, especially those that we tentatively
 classify in emission could not be fitted. Moreover, as we discuss in detail in Section 5, the absorption features do not have all the same blueshift
 velocity and appear to have been produced at different optical depth, making 
 a rigorous fit with the model infeasible. 
 However, because the level and shape
 of the continuum are well reproduced, and so are the strongest absorption features,
 as shown in Fig.~\ref{fig:timecompspecfit4}, the fit allows deriving important physical parameters.
 Although we used three different models of the TMAP grid, 
 varying in nitrogen and carbon abundance, we caution that this does not necessarily
 indicate variation of these elements' abundances mixed in the outer atmospheric
 layer. TMAP models only include the chemical composition of elements of H, He, C, N, O, Ne, Mg, Si, S, Ca and Ni, they lack the
 Ar element, while Ar L-shell ions are rather important between the energy range 20 - 40 \AA \, and two absorption lines (Ar XV, $\lambda_0=28.346$\,\AA; Ar XIII, $\lambda_0=29.497$\,\AA) are identified in the spectra. Only an ad-hoc TMAP model for this nova, with abundances as free parameters and
 including different blueshift velocity for the lines, would
 allow fitting a larger number of lines and obtaining
 a statistically rigorous fit, but it requires extensive modeling
 that is beyond the scope of this observational paper.

For the average spectrum of day 71.3, by leaving all parameters free we obtained 
 a N(H) value of 4.3 $\times$ 10$^{20}$ cm$^{-2}$, but because we could not reproduce the K-edge absorption edge of N VI at 22.46 \AA \ that ``cuts'' the spectrum, we decided to constrain the
 N(H) value at 
 6 $\times$ 10$^{20}$ cm$^{-2}$, close to that obtained
 for the remaining epochs and to the column density in the direction of the nova indicated
 by \citet{2016A&A...594A.116H}, without even assuming any intrinsic absorption of the ejecta. 
Fig.~\ref{fig:timecompspecfit4} and Table~\ref{table:parameters TMAP}
 also show the spectral fit for the first exposure. 
The fit improved when we split the first observation into periods of "high" and "low" count rates,
 but we still had to constrain N(H) to reproduce the absorption edge.
 Most notably, for the first spectrum we could not explain excessive emission at
 23-24 \AA. 

 The best fit value of T$_{\rm eff}$ of day 158.8,
$\sim$805,000 K implies a WD mass in the
 range 1.15-1.25 M$_\odot$ according to
 \citet{Yaron2005}, about 1.25 M$_\odot$ \citep[following][]{2012BaltA..21...76S},
 or a little less than 1.2 M$_\odot$ \citep[according to][]{2013ApJ...777..136W}. 
This is consistent also with the range 
 $1.1~M_\odot \leq M_{\mathrm{WD}} \leq 1.3~M_\odot$ estimated by \citet{Jurdana2012}
 considering the lower limit on the recurrence time and on estimate of mass
 accretion rate $\dot m$.

The strongest absorption features are as deep as predicted by the TMAP model, 
 and they are almost all blue-shifted
 (except the ISM features: O I, N I and C II), however, we show
 in the next Section that the blue shift velocity varies, but we only had uniform blue-shift velocity in the TMAP model fit.
 Moreover, some adjacent absorption lines are blended, complicating the fitting
 procedure. 
 We considered the possibility that two models
 at different temperature and blue-shift velocity
 are needed for the fit, implying a non-uniform WD atmosphere with two different zones,
but this attempt did not improve the fit. 

 These fits predict less flux ($\simeq$25\% for the first
 observation, $\simeq$5\% for the others)
 than measured, indicating that one additional component is
 contributing to the flux. In other novae, this has been attributed to the
 shocked optically thin plasma of the ejecta \citep[.e.g. N LMC 2009,][]{Bode2016}.
In the first observation, clearly the optically
 thin plasma must be contributing more to the total flux budget.
 Values of X-ray luminosity
 of the ejecta of the order of 10$^{34}$ erg s$^{-1}$ are rather typical for
 non-symbiotic novae \citep{Orio2012}.

\begin{figure*}
\begin{center}
\includegraphics[width=8.2cm,clip]{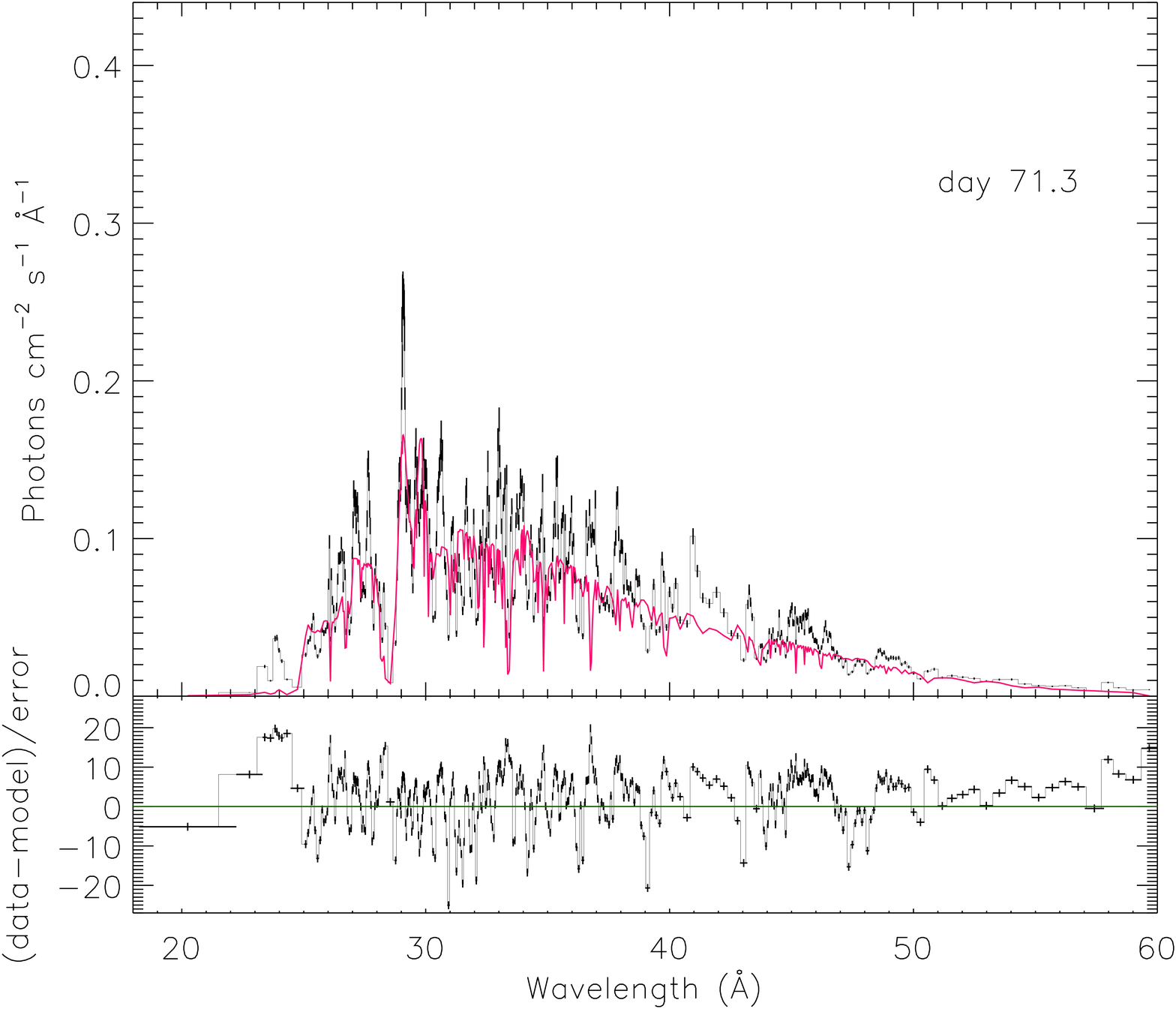}
\hspace{0.22cm}
\includegraphics[width=8.2cm,clip]{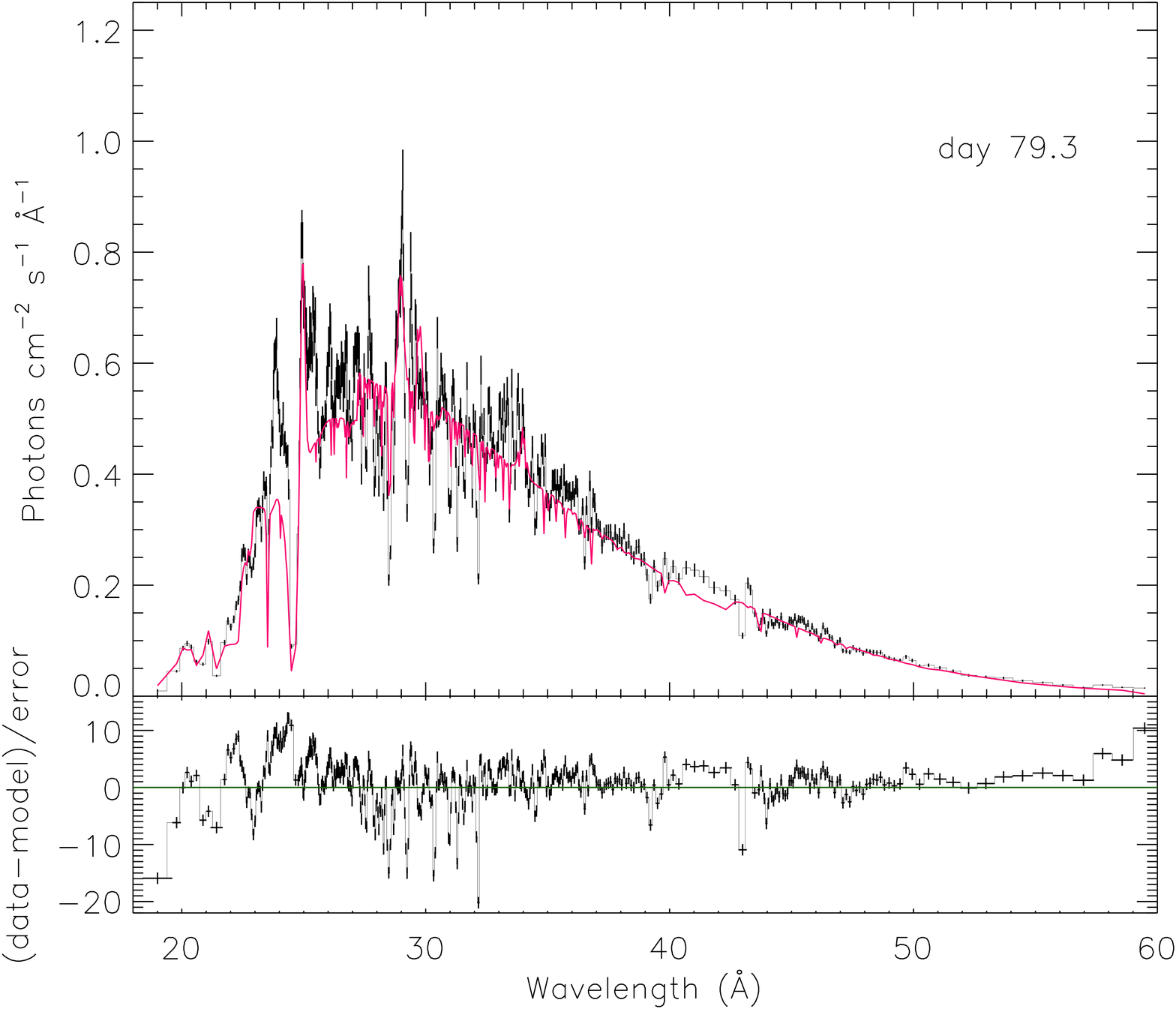}
\caption{The spectra of days 71.3 and 79.3 are 
shown in each upper panel and the fits obtained 
 using TBABS, TMAP with log(g)=9, and BVAPEC are traced with red solid lines. Two of TMAP models 201 and 003 were used for the spectra of days 71.3 and 79.3, respectively. The corresponding residuals of each fit are shown in each lower panel. 
The parameters are reported in Table~\ref{table:TMAP BVAPEC}. Note that, for the first observation, the fit is only
 moderately improved in the range 20$-$30 \AA .}\label{fig: BVAPECfits}
\end{center}
\end{figure*}

In order to fit the additional component, and especially
the emission lines in the range 20$-$30 \AA \ with a wide range red-shifted velocity,
 we added a model of shocked plasma in collisional ionization
 equilibrium, that takes line shifts and line broadening
 into account, namely BVAPEC in XSPEC \citep{2001ApJ...556L..91S}.
 Such a composite model had been found could fit the spectra of novae U Scorpii 
\citep{2013MNRAS.429.1342O}, T Pyxidis \citep{2013ApJ...779...22T}, and V959 Mon \citep{2016ApJ...829....2P}. 
In Fig.~\ref{fig: BVAPECfits} we show the composite model fit for the first two observations,
 and the parameters are reported in Table~\ref{table:TMAP BVAPEC}.
 The fit is improved with the introduction
 a few strong emission features, but several other features are still
 unaccounted for. For the first observation the excess flux at the short
 wavelengths is better modeled by a complex of N VI and N VII lines, but the continuum level is not
 sufficiently high yet at these wavelengths.

 The static model indicates that the peak luminosity (4.90 $^{+0.23}_{-0.18}$ $\times 10^{37}$ erg s$^{-1}$.
 Assuming the unabsorbed flux of the last exposure's spectral fit and the 
distance of $3.69_{-0.42}^{+0.53}$ kpc), the derived absolute
 luminosity is only 
 a fraction of (less than 40\%) the total 
 X-ray luminosity of the WD. In fact, detailed calculations by \citet{Yaron2005}
 show that the bolometric luminosity (of which the X-ray luminosity at
 T$_{\rm eff} > 500,000$ K constitutes more than 95\%),
 for a WD mass of at least 1 M$_\odot$ it would always exceed 1.3 $ \times 10^{38}$ erg s$^{-1}$.

\begin{figure*}
\begin{center}
\includegraphics[width=8.9cm]{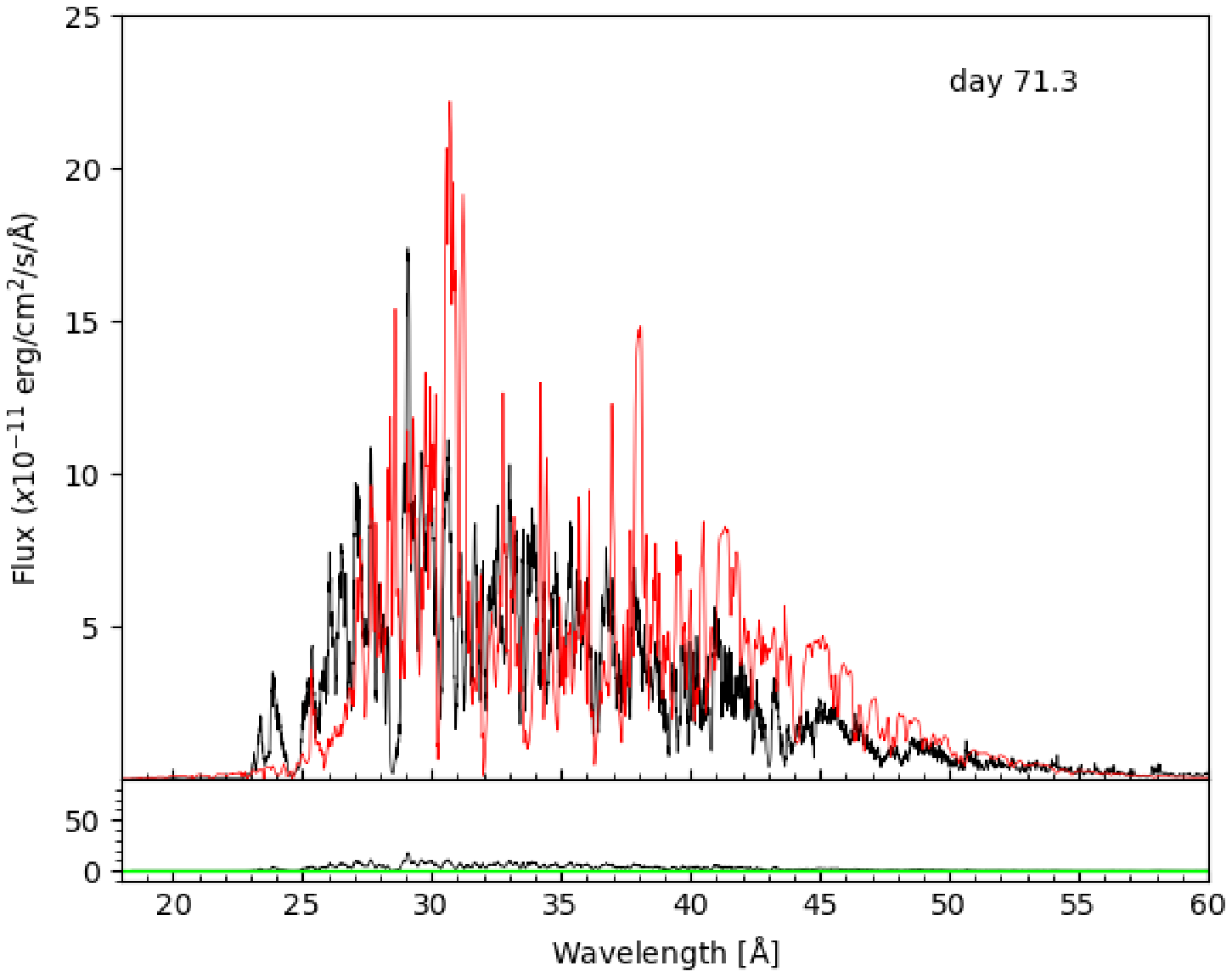}
\hspace{-0.82em}
\includegraphics[width=8.9cm]{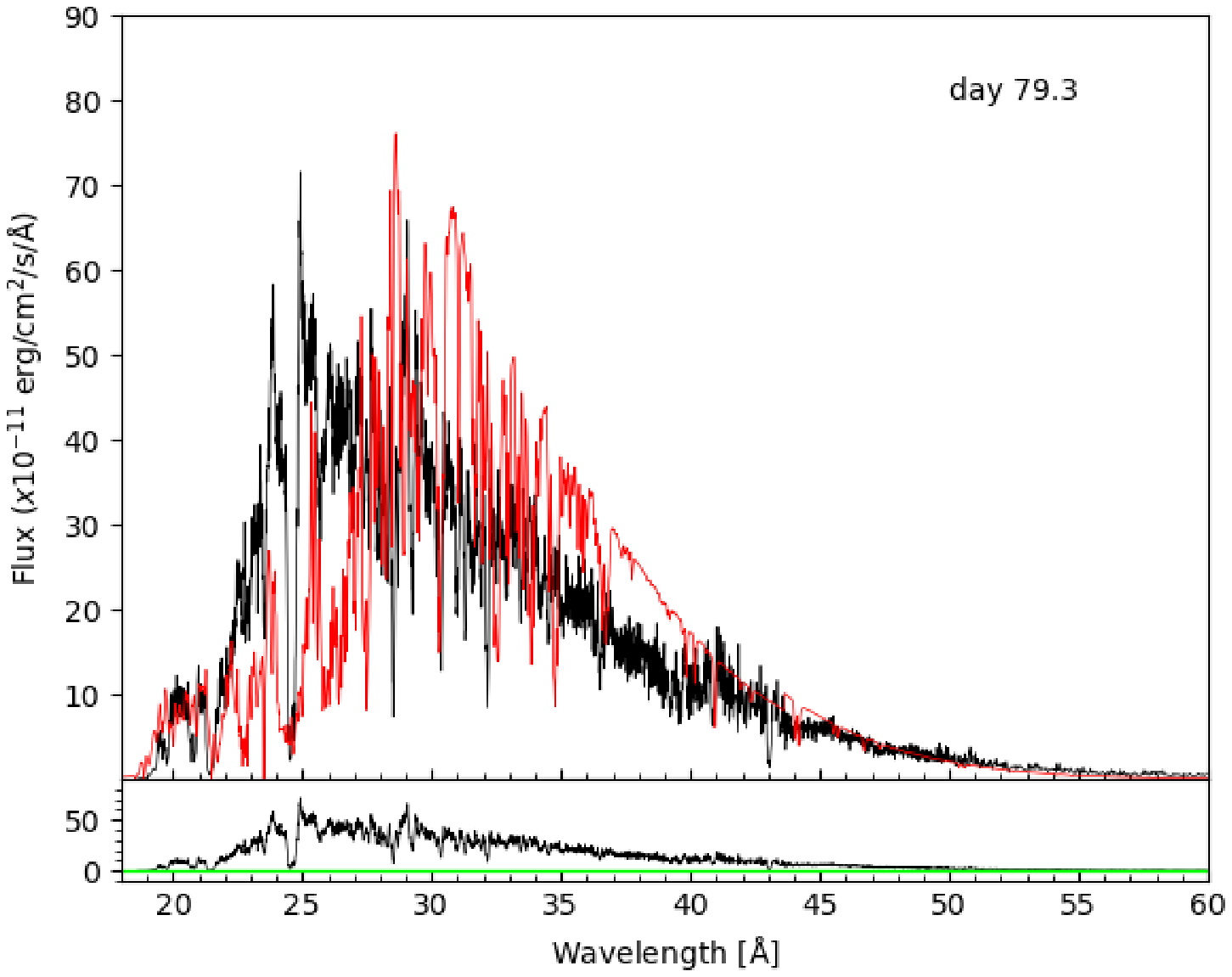}
\hspace{-0.82em}
\includegraphics[width=8.9cm]{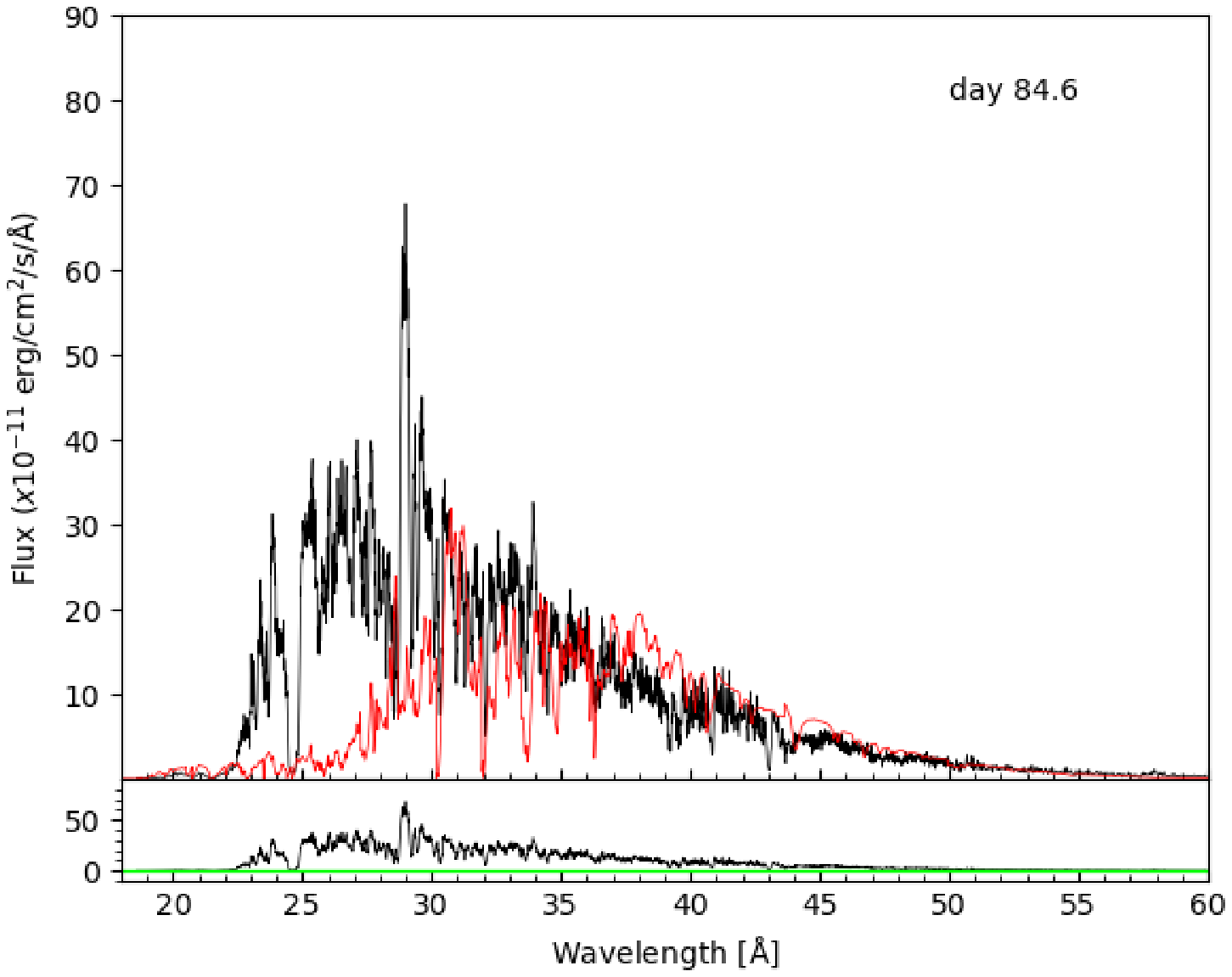}
\hspace{-0.82em}
\includegraphics[width=8.9cm]{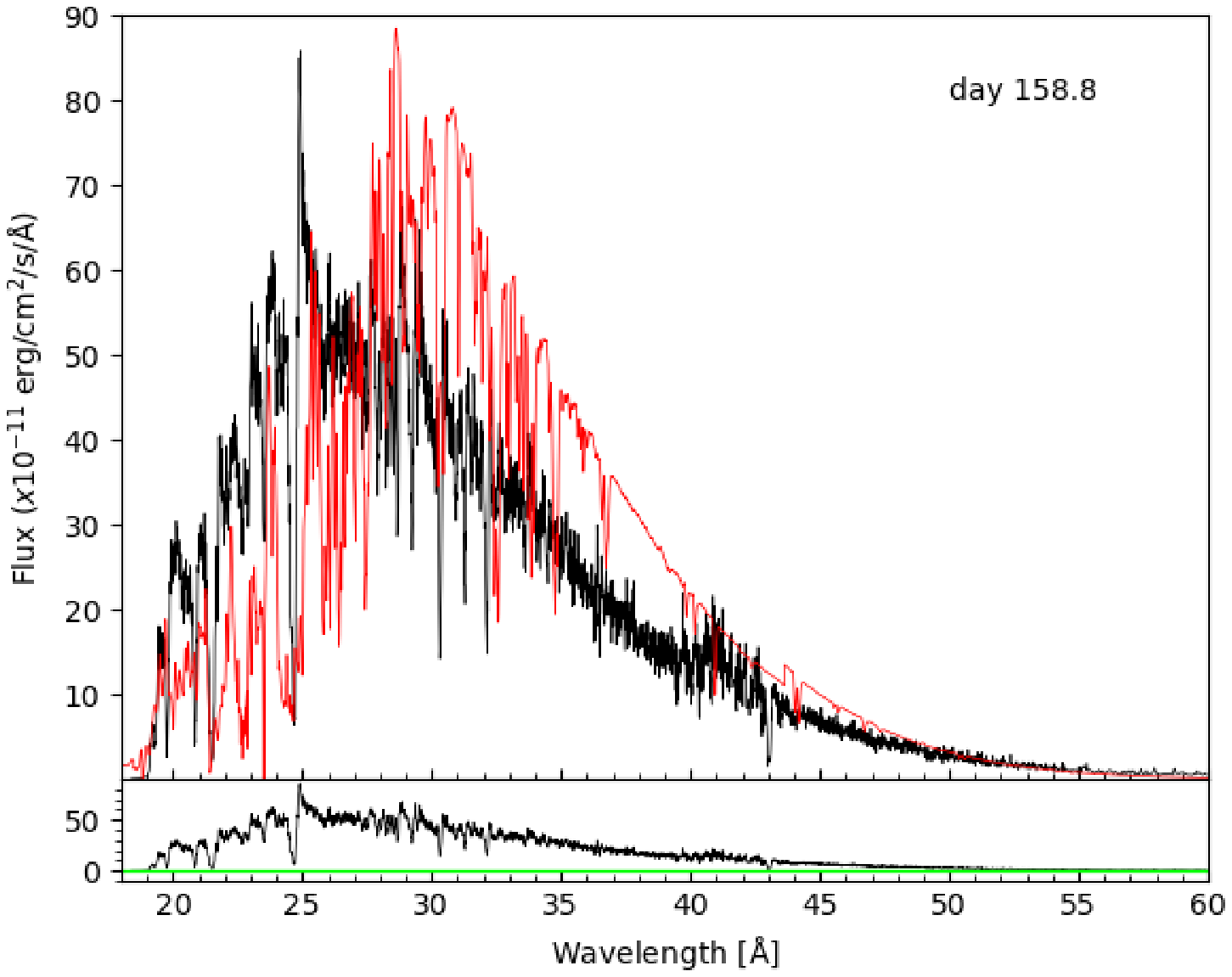}
\caption{Each upper panel shows
 the spectra (in black) and the fits obtained using the WT model (in red). 
The corresponding residuals of each fit are shown in each lower panel. The parameters are reported in Table~\ref{table:parameters wind}.}\label{fig: spectra van}
\end{center}
\end{figure*}
\begin{table*}
\caption{Parameters of the best fit obtained with
 TMAP. The unabsorbed flux is calculated in the 10-124 \AA \ range \ ($\approx$0.10-1.24 keV). 
N(H) has a fixed value 6.00 $\times$ 10$^{20}$ cm$^{-2}$ for the fits of the day 71.3 spectra.
 The nitrogen and carbon abundance with respect to the solar value are not resulting
 from the fit. There
 is a discrete grid of abundances in the TMAP public models and this is the set of abundances that yields the best fit. 
The uncertainty on the value of N(H) and the absorbed
 and unabsorbed flux are calculated by fixing all the other parameters.
 The temperature grid is interpolated
 with steps of 10,000 K, so this is the realistic value of the uncertainty. The errors are at a 90 percent confidence level.}
\label{table:parameters TMAP}
\begin{center}
\begin{tabular}{rrrrrrrr}\hline\hline \noalign{\smallskip}
Day & T$_{\rm eff}$ (K) & N(H) (10$^{20}$ cm$^{-2}$) & Velocity (km s$^{-1}$)$^a$ & F(abs) $\times 10^{-9}$ &
F(unabs) $\times 10^{-9}$ & [N/N$_\odot$]$^b$ & [C/C$_\odot$]$^b$  \\ 
    &                   &                     &       & (erg cm$^{-2}$ s$^{-1}$) & (erg cm$^{-2}$ s$^{-1}$) &  & \\
\hline
71.3 & 566,430  & 6.00 &  3029$^{+17}_{-16}$ & 0.71 $^{+0.01}_{-0.01}$ & 6.01$^{+0.02}_{-0.02}$  & 1.668 & -0.057 \\
 71.3 (High)$^c$ & 571,493  & 6.00 & 3029$^{+22}_{-23}$   & 0.97$^{+0.01}_{-0.01}$ & 8.38$^{+0.03}_{-0.04}$ & 1.668 & -0.057  \\
 71.3 (Low)$^d$ &  559,860  & 6.00 &  3029$^{+18}_{-25}$  & 0.44$^{+0.01}_{-0.01}$ & 4.33$^{+0.02}_{-0.02}$ & 1.668 & -0.057  \\
 79.3 & 760,344  &  5.73$^{+0.04}_{-0.03}$  & 2771$^{+74}_{-57}$   &  5.48 $^{+0.02}_{-0.02}$ & 23.95$^{+0.16}_{-0.19}$ & 1.803 & -1.513 \\
 84.6 & 721,063  & 5.61$^{+0.04}_{-0.02}$   &  1213$^{+20}_{-17}$  & 3.44 $^{+0.01}_{-0.01}$ & 16.48$^{+0.18}_{-0.11}$ & 1.803 & -1.513   \\
 158.8 & 805,023  & 5.74 $^{+0.04}_{-0.04}$ & 1126$^{+15}_{-16}$   & 7.55 $^{+0.02}_{-0.02}$ & 30.12$^{+0.20}_{-0.18}$ & 1.678 & -1.073  \\
\hline  \noalign{\smallskip}
\end{tabular}
\end{center}
Notes:\hspace{0.1cm} $^a $: The velocity of TMAP. $^b $: Logarithm of abundance ratio. 
$^c $: Spectrum extracted for the
 period during which the zero order light curve count rate was larger than the average 13.87 counts s$^{-1}$. $^d $: Spectrum extracted for the period
 during which the zero order light curve count rate was lower than the average 13.87 counts s$^{-1}$.\\
\end{table*}
\begin{table*}
\caption{Parameters of the best fits obtained with TMAP+BVAPEC. For the atmosphere, the nitrogen and
 carbon abundance are the same as in the model above (Table~\ref{table:parameters TMAP}). The BVAPEC
absorbed flux is about 10\% of the
 values in Table~\ref{table:parameters TMAP}. However,
 BVAPEC contributes to 5\% of the unabsorbed flux in the first observation and to 2\% in the second.}
\label{table:TMAP BVAPEC}
\begin{center}
\begin{tabular}{rrrrrrrrrr}\hline\hline \noalign{\smallskip}
Day & T$_{\rm eff}$ (K) & N(H) (10$^{20}$ cm$^{-2}$) & V$_{\rm bs}$$^a$  (km s$^{-1}$) & T$_{\rm bv}$ (eV)$^b$ & V$_{\rm rs}$  (km s$^{-1}$)$^c$ & v$_{\rm b}$ (km s$^{-1}$)$^d$ & [N/N$_\odot$] & [C/C$_\odot$] &  [S/S$_\odot$] \\
\hline
71.3 & 560,517 & 6 (fixed)   & 2999$^{+22}_{-26}$ & 81  & 2794$^{+31}_{-94}$ & 1500$^{+48}_{-48}$ & 1000 & 73 & 143 \\
79.3 & 758,894 & 5.61$\pm04$ & 2737$^{+73}_{-78}$ & 150$^{+26}_{-43}$ & 2025$^{+27}_{-59}$ & 1074$^{+35}_{-34}$ & 1000 & 152  & 0.003 \\  
\hline  \noalign{\smallskip}
\end{tabular}
\end{center}
Notes:\hspace{0.1cm} $^a $: The velocity of TMAP. $^b $: The temperature of BVAPEC. $^c $: The centroid shift of BVAPEC. $^d $: The broadening velocity of BVAPEC.\\

\end{table*}
\subsection{Fitting the spectrum with the ``wind atmosphere'' model}
The other grid of models we used was
obtained with the ``wind-type'' (WT) PHOENIX expanding atmosphere code, which
includes a larger number of atomic species, but was developed only with solar
abundances \citep{vanRossum2012}. Since abundances are very different from
 solar in the burning layer and burning ashes, and convection mixes 
 this material abundantly in the WD atmosphere, this is of course
 a drawback in using the public model grid. 
The model assumes a hydrostatic base, parameterized with effective
 temperature T$_{\rm eff}$, radius R and gravitational acceleration log(g), 
 and an envelope with a constant mass loss rate $\dot M$ and a velocity v, with law
 similar to that of the winds of massive stars, namely

 \begin{equation}
 v(r>R) = v_0 +(v_{\infty}-v_0) \left(1-\frac{R}{r}\right)^{\beta}
 \end{equation}

 where $\beta$=1.2 in order to have a smooth transition between the static core and the 
 expanding envelope, and the asymptotic
 velocity v$_\infty$ is the fifth parameter of the public calculated grid.
 The values of $\dot M$ are between 10$^{-7}$ and 10$^{-9}$ M$_\odot$ yr$^{-1}$, thus
 the mass loss is assumed to be quite small compared to the values
 during the nova outflows before the supersoft X-ray phase.
 At higher $\dot M$ the lines are not
 just in absorption, but they have emission wings 
 and a P-Cyg profile forms. On the other hand, 
 the higher the temperature, the less prominent are these emission
 wings, and at the temperature of the novae observed so far, not many and
 not very strong emission wings 
 are not produced in the atmosphere \citep[see][]{Orio2018}. In addition, the large blue-shift velocities observed so far,
 exceeding 1000 km s$^{-1}$, tend to smooth these P-Cyg profiles and
 make them difficult to detect and measure.
 In other words, we do not expect the WT model to explain the emission
 lines of the X-ray spectra of novae as formed in the atmosphere or in
 the wind close to the WD. They are more likely to be due to shocks
 and to form far from the WD surface \citep[see also][]{2013A&A...559A..50N}.

 Other characteristics of the model are the much less prominent and smoother
 absorption edges, and because the configuration
 is not static, the effective gravity can be lower (due
 to a larger radius), since the outflow implies that 
 balancing the radiation pressure is not required. Thus, g 
 is smaller than in the static models with the same T$_{\rm eff}$. There is also very
 little dependence of the spectrum on the WD mass, because T$_{\rm eff}$ is not a function
 of the mass as in the static models: the WD radius varies and is much larger than in
 the static configuration. Thus, it is interesting to experiment and
 try and fit the spectra also with this model, to see whether it better
 fits the continuum and the absorption features, accounting for their blue-shift. 

 We fitted the fluxed spectra (corrected for effective area)
 with the models of the WT grid in IDL. We did not attempt to choose a best fitting model
 by minimizing statistical parameter. Most models
 in the grid are inadequate and were ruled out immediately, since they predict 
high flux in a much broader wavelength range. 
In order to take the intervening column density into account,
we used again the T\"ubingen-Boulder absorption model and
performed the calculation with the related routine TBNEW in IDL.

 Table~\ref{table:parameters wind} reports the parameters of the grid we chose because they 
 fit better than others the continuum shape and the absorption edge causing abrupt flux decrease at
 the higher energy. The results are shown
 in Fig.~\ref{fig: spectra van}. Because the high blue shift velocity smooths the absorption edges and
 lines excessively for this spectrum, we compensated by assuming a higher value of the absorbing
 column density, which would indicate some amount of intrinsic absorption in the ejecta. However,
 the N(H) value causes too little flux below 50-55 \AA. We could not solve this problem
 and fit the hard and soft portion of the spectrum simultaneously. 
 As expected, T$_{\rm eff}$ is lower than that resulting
 from the TMAP best fit, and $\dot M$ is of the order of 10$^{-8}$ M$_\odot$ yr$^{-1}$.
 Altogether, TMAP 
 fits the continuum and the strongest absorption
 lines much better than this model. We have interpolated
 between the tabulated models from van Rossum's archive\footnote{http://flash.uchicago.edu/\~{}daan/WT\_SSS/}, which were developed specifically for another nova (V4743 Sgr), so
 we were not expecting a perfect fit, but the large discrepancy between this model
 and observed spectra is disappointing. 
 We note that, with larger absorbing column density N(H), in this model the unabsorbed flux in
 the last exposure reaches 7 $\times 10^{-8}$ 
 erg cm$^{-2}$ s$^{-1}$, implying that a larger portion (less than 85\%) of the WD was 
visible to us. However, the model shown in the figure
 overestimates the absorbed flux in the X-ray band
 (see comparison with Table~\ref{table:obs}).
\begin{table*}
\caption{Parameters of the fit using the model of \citet{vanRossum2012} shown in Fig.~\ref{fig: spectra van}. To normalize the flux, a distance of 3.69 kpc was assumed.}
\label{table:parameters wind}
\begin{center}
\begin{tabular}{rrrrrrr}\hline\hline \noalign{\smallskip}
Day & T$_{\rm eff}$ (K) & N(H) (10$^{20}$ cm$^{-2}$) & Velocity (km s$^{-1}$) & log(g) (cm s$^{-2}$)&
 $\dot m$ (M$_\odot$ year$^{-1}$) & F(abs) (erg cm$^{-2}$ s$^{-1}$) \\ 
\hline
71.3 &  475,000 & 9.70 & 2400 &  8.49  & 1.0  $\times 10^{-8}$ & 1.19 $\times 10^{-9}$ \\
 79.3 & 500,000 & 9.50  & 1800 &  7.77  & 7.6  $\times 10^{-9}$ & 6.21 $\times 10^{-9}$ \\
 84.6 & 475,000 & 8.50  & 2400  &  8.32  & 3.1  $\times 10^{-8}$ & 2.53 $\times 10^{-9}$  \\
 158.8 & 500,000 & 8.50 & 2400 &  7.73  & 1.0  $\times 10^{-8}$ & 8.81 $\times 10^{-9}$ \\
\hline  \noalign{\smallskip}
\end{tabular}
\end{center}
\end{table*}
\section{Discussion }
\label{sec:discussion}
 The X-ray luminosity of KT Eri and its variation is consistent with the possibility that we observed only a fraction of the WD surface, 
 with visibility varying within hours.
This seems to be a common phenomenon in many novae, and was observed,
 for instance, also in N LMC 2009 \citep{Orio2021} and
 N SMC 2016 \citep{Orio2018}. 
 Detailed calculations show that the thermonuclear
 runaway always expands all over the surface and is nearly spherically symmetric 
\citep{Glasner1997}, so the supersoft flux is also expected to be homogeneous in
the WD surface, at least until the time magnetically driven accretion onto the WD
 is resumed on strongly magnetized WDs \citep[see][]{Aydi2018, Zemko2018}. 
The likely explanation we propose for KT Eri is 
that there were dense clumps in the ejecta,
 which were optically thick to the X-rays. Although it may seem more likely to produce a clumpy outflow in the
 initial explosive emission, the SSS variation in KT Eri was observed so long after
 the outburst maximum that initial clumps must have been ejected far away from
 the WD, and be located where the WD would appear as a point source, so partial
 obscuration by these initial ejecta would be unlikely. We suggest that the 
 optically thick radiation driven wind that is supposed to last for months
 \citep[initially suggested by][and always included in the nova models]{Bath1976}
 is not a continuous and 
 smooth phenomenon. Even if this variable wind has never been modeled,
 many data can be explained with 
 distinct ejection episodes at different
 wind velocity, including observations of novae T Pyx \citep{Chomiuk2014}
 and V906 Car \citep{Sokolovsky2020}. Distinct outflows may be a common occurrence 
also according to \citet{Aydi2020}.
 Instabilities are likely to occur in such a ``stunted'' outflow, and may cause sufficiently small scale clumpiness to obscure only a portion of the WD, so that the filling factor of
 the ejecta along the line of sight varied within hours.

 The X-ray grating spectra allow to draw some conclusions
 regarding the WD mass. \citet{vanRossum2012}
 shows that the effect of an ongoing outflow is to lower the atmospheric temperature,
 and somewhat bloat the radius, reaching the same luminosity of
 a static configuration at lower T$_{\rm eff}$.
However, because the full fledged evolutionary models of the outburst
 (e.g., MESA used by \citet{2013ApJ...777..136W}, or the model
 calculated by \citet{Yaron2005} and references therein) also assume a static atmosphere, from which mass loss has ceased
 at the peak of the supersoft X-ray phase, the results of the outburst
models can
 be compared with the parameters derived with TMAP, which
 is also static. For KT Eri, from the TMAP fit
we obtain a peak T$_{\rm eff}$ of about
 800,000 K, corresponding to a WD mass in the range 1.15 $-$ 1.25 M$_\odot$ according to all nova outburst models quoted above.
It is interesting to notice
 that \citet{Jurdana2012} estimated an approximate pre-outburst
 accretion rate $\dot m$ from the optical
 B luminosity at quiescence, and by
 assuming a recurrence time of more than 100 years (because
 no previous outburst was found in the Harvard plates),
these authors reached the same conclusion
 concerning the WD mass of KT Eri.

The above result is in agreement with the parameters derived
correlating also additional
observational evidence with the models. The {\sl Swift} XRT light curve shows that the
 SSS flux decreased by a factor close to 16 in exactly 7 months after optical maximum.
 210 days is thus the
 duration of the ``bolometric t$_3$'' (time for a decrease by
 the equivalent of 3 mag in optical),
 which is the parameter adopted by \citet{Yaron2005} to evaluate
 the duration
 of the constant bolometric luminosity phase of a nova.
 This parameter turns out to be of the
 order of 200 days for a 1.25 M$_\odot$ WD for a large range of initial WD surface temperature and
 for a pre-outburst mass accretion rate $\dot m > 10^{-11}$ M$_\odot$ yr$^{-1}$,
 while a 1 M$_\odot$ WD nova with
 the same ``bolometric t$_3$''
would have been initially cool, and would have undergone mass
 transfer in a narrow range $10^{-10}$ M$_\odot$ yr$^{-1} < \dot m < 10^{-9}$ M$_\odot$ yr$^{-1}$. We now know also the X-ray luminosity of
 KT Eri at quiescence: assuming that the X-rays
 originate in the boundary
 layer of a geometrically thin accretion disk, \citet{Sun2020}
indeed derived a consistent value, $\dot m = 2 \times 10^{-10}$
 M$_\odot$ yr$^{-1}$ 9 years after the outburst, which 
 can be regarded as a lower limit, because it 
 may be higher if there is an accretion disk corona emitting only in the UV.

\section{Conclusions}
\label{sec:conclusions}
 The spectra of novae in the X-ray supersoft phase are complex,
 but they contain a wealth of physical information.
 We found that the KT Eri spectrum has a range of blue shift
 velocity of the absorption lines and red shift velocity of the emission lines. By fitting
 models with a single velocity, the fit is driven by the strongest features.
 A future, comprehensive model should 
 account also for absorption features produced in
 layers of different optical depth and velocity. We calculated the velocity of
 the strongest absorption features and optical depth at which they are produced,
 as useful data to study more complex physical models in the near future. 

 The analysis of the X-ray spectra of KT Eri
is complicated by the large irregular variability in two of the exposures,
 and by the overlapping emission lines,
 which we attributed to the ejecta as suggested for most
 novae and SSS by \citet{Ness2013}.
 These emission lines are not predicted to originate from, or near, the WD by 
 the static atmospheric models, neither by the wind-atmosphere
 models. In the static atmosphere, conspicuous
emission lines are formed
 only for the lowest WD masses \citep{2010ApJ...717..363R}, at 
 much lower T$_{\rm eff}$ than that resulting from the KT Eri spectra.
 The wind-atmosphere model of \citet{vanRossum2012} instead
 predicts more emission lines and actual P-Cyg profiles, but
 they are smoothed and even ``washed out'' by the wind velocity.
We could not model the emission line spectrum by adding one or two
 CIE components, and concluded that
 it is produced in a complex emission region,
 and perhaps it is not only due to shock ionization.
 In at least two exposures, the variation of the flux in some emission lines
is independent of the variations of the continuum SSS flux and
 may be related by variations in the neutral oxygen column density. We suggested 
 that either the surrounding cool material is not homogeneous, or
 that it was temporarily ionized. 

 Our analysis of these spectra was phenomenological and included the comparison with
 atmospheric, wind and outburst models. We were able to conclude that the nova WD
 has mass in the 1.15-1.25 M$_\odot$ range and that the mass accretion rate $\dot m$ 
before the outburst was most likely close to a value
 of 10$^{-10}$ M$_\odot$ yr$^{-1}$. 

Finally, we also confirmed
 that pulsations of the SSS with a $\simeq$35 s timescale appeared only late in
 the post-outburst X-ray light curve, more than 100 days after optical maximum. 
 We note that this was quite different from the case of RS Oph, in which a $\simeq$35 s periodicity 
 appeared only in the first $\simeq$55 days after optical maximum \citep[see][]{Nelson2008,
Osborne2011, Ness2013}, We suggest that the search for the root cause of this phenomenon
 will have to make use of this fact and correlate it with other
differences in the physical parameters of the two novae.
 
\section*{Data Availability}
The data analyzed in this article are all available in the HEASARC archive of NASA at the following URL: \url{https://heasarc.gsfc.nasa.gov/db-perl/W3Browse/w3browse.pl}

\section*{Acknowledgements}
We thank the anonymous referee for her or his constructive comments and suggestions, which helped us to improve the scientific content of this paper. S. Pei has been supported by the China Scholarship Council (No. 201704910918)
 and M. Orio by a Chandra-Smithsonian grant for archival research at the University 
 of Wisconsin.




\bibliographystyle{mnras}
\bibliography{KTEri} 




\bsp	
\label{lastpage}
\end{document}